\documentclass[journal]{IEEEtran}

\usepackage{mathrsfs}
\usepackage[noadjust]{cite}
\usepackage{graphicx,color,overpic,psfrag}
\usepackage{amsmath, amssymb}
\usepackage{latexsym}
\usepackage{bm}
\usepackage{amssymb}
\usepackage{cases}
\usepackage{array}
\usepackage{fancyhdr}
\usepackage{setspace}

\ifCLASSOPTIONcompsoc
\usepackage[caption=false,font=normalsize,labelfont=sf,textfont=sf]{subfig}
\else
\usepackage[caption=false,font=footnotesize]{subfig}
\fi

\usepackage{url}
\usepackage{algpseudocode}
\usepackage{algorithm}
\usepackage{blkarray}
\usepackage{booktabs}

\usepackage{multirow}
\usepackage{dsfont}
\usepackage{tabularx}
\usepackage[table]{xcolor}

\usepackage{amsfonts}

\usepackage{letltxmacro}

\graphicspath{{figure/}}%path of the figures

\newtheorem{prop}{Proposition}

\newcommand{\figref}[1]{Fig. \ref{#1}}

\newcommand{\alref}[1]{Algorithm \ref{#1}}
\newcommand{\appref}[1]{Appendix \ref{#1}}
\newcommand{\secref}[1]{Section \ref{#1}}

\newcommand{\propref}[1]{{Proposition \ref{#1}}}
\newcommand{\tr}[1]{\mathsf{tr}\left\{#1\right\}}
\newcommand{\diag}[1]{\mathsf{diag}\left\{#1\right\}}

\newcommand{\argmax}[1]{\mathop{\arg\max}\limits_{#1}}

\newcommand{\cP}{\mathcal{P}}

\newcommand{\ba}{\mathbf{a}}

\newcommand{\bn}{\mathbf{n}}

\newcommand{\bs}{\mathbf{s}}

\newcommand{\bx}{\mathbf{x}}
\newcommand{\by}{\mathbf{y}}

\newcommand{\bA}{\mathbf{A}}
\newcommand{\bB}{\mathbf{B}}
\newcommand{\bC}{\mathbf{C}}

\newcommand{\bF}{\mathbf{F}}

\newcommand{\bH}{\mathbf{H}}
\newcommand{\bI}{\mathbf{I}}

\newcommand{\bK}{\mathbf{K}}

\newcommand{\bQ}{\mathbf{Q}}
\newcommand{\bR}{\mathbf{R}}
\newcommand{\bS}{\mathbf{S}}

\newcommand{\bU}{\mathbf{U}}
\newcommand{\bV}{\mathbf{V}}
\newcommand{\bW}{\mathbf{W}}
\newcommand{\bX}{\mathbf{X}}

\newcommand{\bSigma}{{\boldsymbol\Sigma}}

\newcommand{\bLambda}{{\boldsymbol\Lambda}}
\newcommand{\bOmega}{{\boldsymbol\Omega}}

\newcommand{\bPi}{{\boldsymbol\Pi}}

\newcommand{\ntb}{\notag\\}

\newcolumntype{L}{>{\hspace*{-\tabcolsep}}l}
\newcolumntype{R}{c<{\hspace*{-\tabcolsep}}}
\definecolor{lightblue}{rgb}{0.93,0.95,1.0}

\newcommand{\C}{\mathbb{C}}

\newcommand{\Omegak}{\mathbf{\Omega}_{k}}

\newcommand{\Pmaxk}{P_{\mathrm{max},k}}
\newcommand{\Pmax}{P_{\mathrm{max}}}

\newcommand{\etaSE}{{\eta}_{\mathrm{SE}}}

\newcommand{\etaEE}{{\eta}_{\mathrm{EE}}}

\newcommand{\etaEEb}{{\overline {\eta}}_{\mathrm{EE}}}

\newcommand{\sigmatwo}{\sigma^{2}}

\newcommand{\bGamma}{\bm{\Gamma}}

\newcommand{\bPhi}{\bm{\Phi}}

\newcommand{\bDelta}{\bm{\Delta}}

\newcommand{\bpi}{\bm{\pi}}
\newcommand{\wbGamma}{\widetilde{\bm{\Gamma}}}
\newcommand{\wbPhi}{\widetilde{\bm{\Phi}}}

\newcommand{\bTheta}{\bm{\Theta}}
\newcommand{\wbTheta}{\widetilde{\bm{\Theta}}}

\newcommand{\Rka}{\mathbf{R}_{k,a}}
\newcommand{\wbK}{\widetilde{\mathbf{K}}}
\newcommand{\wbH}{\widetilde{\mathbf{H}}}
\allowdisplaybreaks

\begin{document}

%\title{\huge Spectral Efficiency and Energy Efficiency Tradeoff in Uplink MIMO Transmission with Electromagnetic Exposure Constraints}

\title{Rate-Splitting Multiple Access for Uplink Massive MIMO With Electromagnetic Exposure Constraints}

\author{
	Hanyu~Jiang,
	Li~You,
	Ahmed~Elzanaty,
	Jue~Wang,
	Wenjin~Wang,
	Xiqi~Gao,
	and~Mohamed-Slim~Alouini%
\thanks{
	Copyright (c) 2015 IEEE. Personal use of this material is permitted. However, permission to use this material for any other purposes must be obtained from the IEEE by sending a request to pubs-permissions@ieee.org.
}
\thanks{
Part of this work was presented in ICSPS 2022 \cite{ICSPS2022}.
}
\thanks{
	Hanyu Jiang, Li You, Wenjin Wang, and Xiqi Gao are with the National
	Mobile Communications Research Laboratory, Southeast University, Nanjing
	210096, China, and also with the Purple Mountain Laboratories, Nanjing 211100, China (e-mail: hyjiang@seu.edu.cn; lyou@seu.edu.cn; wangwj@seu.edu.cn; xqgao@seu.edu.cn).
	
	Ahmed Elzanaty is with the 5GIC \& 6GIC, Institute for Communication Systems, University of Surrey, Guildford, GU2 7XH, United Kingdom (email: a.elzanaty@surrey.ac.uk).
	
	Jue Wang is with School of Information Science and Technology, Nantong
	University, Nantong 226019, China, and also with Nantong Research Institute for Advanced Communication Technologies, Nantong 226019, China (e-mail: wangjue@ntu.edu.cn).
	
	Mohamed-Slim Alouini is with the Computer Electrical and Mathematical Science and Engineering Division, King Abdullah University of Science and Technology, Thuwal, Makkah Province, Saudi Arabia (e-mail: slim.alouini@kaust.edu.sa).
		
}
}

\maketitle

\begin{abstract}
Over the past few years, the prevalence of wireless devices has become one of the essential sources of electromagnetic (EM) radiation to the public. Facing with the swift development of wireless communications, people are skeptical about the risks of long-term exposure to EM radiation. As EM exposure is required to be restricted at user terminals, it is inefficient to blindly decrease the transmit power, which leads to limited spectral efficiency and energy efficiency (EE). Recently, rate-splitting multiple access (RSMA) has been proposed as an effective way to provide higher wireless transmission performance, which is a promising technology for future wireless communications. To this end, we propose using RSMA to increase the EE of massive MIMO uplink while limiting the EM exposure of users. In particularly, we investigate the optimization of the transmit covariance matrices and decoding order using statistical channel state information (CSI). The problem is formulated as non-convex mixed integer program, which is in general difficult to handle. We first propose a modified water-filling scheme to obtain the transmit covariance matrices with fixed decoding order. Then, a greedy approach is proposed to obtain the decoding permutation. Numerical results verify the effectiveness of the proposed EM exposure-aware EE maximization scheme for uplink RSMA.

\end{abstract}

\begin{IEEEkeywords}
Electromagnetic (EM) exposure, multiuser massive MIMO, rate-splitting multiple access (RSMA), energy efficiency (EE), statistical channel state information (CSI).

\end{IEEEkeywords}
%
%\newpage

\section{Introduction}
Driven by the emerging data-hungry applications, such as virtual reality, cloud core networks, artificial intelligence, and so on, future wireless networks are demanded to support high spectral efficiency (SE) that reaches the peak of thousands of megabits per second, which is usually accompanied by the huge density of wireless connectivity \cite{SBC20}. As a result, there is an emerging concern that people are more likely to be exposed to high doses of electromagnetic (EM) radiation \cite{JHB20,ECA21,JYW22,SBC20}. As indicated in \cite{CEA21}, the biological effects of EM exposure include thermal and non-thermal effects. The former represents the effect that excessive EM radiation absorbed by the body can produce heat that may lead to tissue damage. To deal with this mechanism, regulatory agencies such as the Federal Communications Commission (FCC) and International Commission on Non-Ionizing Radiation Protection (ICNIRP) stipulate the maximum user exposure of EM emitted by a qualified wireless device to ensure that the thermal effects is below some thresholds \cite{CEA21}. In addition, there is a debate about the long-term non-thermal effects on the human body, which requires further research \cite{CEA21}. Since users are mainly exposed to the radiation emitted by wireless terminals, the uplink EM exposure is usually more essential than the downlink \cite{JHB20}. In uplink transmissions, EM exposure is measured by a standard metric named specific absorption rate (SAR), which indicates the EM power absorbed by human tissue per unit mass \cite{JHB20}. Compared with the constraint on power, which is only related to the amplitudes of multiple antennas, SAR is also the function of phase differences between any two transmit signals \cite{HLY13,HB22}. The SAR model is based on experimental studies, which can be characterized by the sinusoidal function of the phase difference \cite{CCM04}. In single antenna cases, the worst-case SAR can be naturally satisfied by cutting down the transmit power. However,  adopting the same approach for the compliance of SAR is not effective in achieving high data rate when there are multiple transmit antennas. This leads to the practical demand for the EM exposure-aware precoding design in multiuser massive multiple-input multiple-output (MIMO) uplink transmissions. 

The introduction of SAR constraints makes it risky to blindly increase the transmit power for higher performance as the power consumption and EM radiation may also surge in excess \cite{JYW22}. Achieving high data rate with low energy consumption and EM exposure has become a significant need that the new generation of wireless communications has to address \cite{JHB20,CEA21}. With this in mind, rate-splitting multiple access (RSMA) is exploited with great potential for MIMO transmission in the next-generation wireless networks to meet the multiplying of mobile data streams and limited resources \cite{YCS22}. Similar to non-orthogonal multiple access (NOMA), which allows multiple users to access the same time and frequency resource block, RSMA further splits the transmit signal of each user into various sub-signals, which are transmitted independently on the corresponding layers \cite{YCS22,ZLN19,MCL19,MMDC22}. At the base station (BS), the receiver adopts successive interference cancellation (SIC) technology for decoding, where the sub-signals of various users and layers are allocated with different rates according to the decoding order \cite{KJU21,XCC21}. 

Compared with conventional multiple access schemes, RSMA makes full use of the available time, frequency, power, and space domains to achieve larger degree of freedom and capacity region in the uplink transmission without time sharing between users \cite{MMDC22}.  
In practical systems, the SE improvement of RSMA compared with NOMA mainly lies in that the length of the decoding order permutation in RSMA is larger than that in NOMA, which increases the degrees of freedom. This property also improves the interference management ability of communications and thus allows RSMA to provide a more robust transmission scheme. 

Recent studies have shown that the quality of communications can be further improved by using RSMA in both downlink and uplink transmissions. In \cite{KJU21}, one-layer rate-splitting was used in the downlink MIMO broadcast channels to optimize the weighted sum rate. The work in \cite{XCC21} investigated the implementation of RSMA in downlink dual-functional radar communications. In \cite{KS21}, a successive null-space-based precoding method was proposed to reduce the inter-user interference and optimize the weighted sum rate of the downlink MIMO-RSMA transmission. In fact, most studies such as \cite{KJU21,XCC21,KS21} focus on the downlink RSMA, where the rate is split by dividing the transmit signal into the common and private parts. For uplink RSMA communication, the capacity can be theoretically achieved with perfect channel state information (CSI) \cite{RU96}. The authors in \cite{ZLN19} proposed a two-layer rate-splitting scheme to guarantee the max-min fairness of single-input multiple-output (SIMO) uplink. In addition, \cite{YCS22} investigated the problem of maximizing the sum rate of users with proportional rate constraints in uplink RSMA transmission, where the message of each user is split into two layers that are transmitted in the same time-frequency resource. Note that the prior works on uplink RSMA mainly focus on the low-complexity precoding methods for the two-layer rate-splitting scheme. These approaches are inapplicable to the scene where the data stream of each user is split into more layers to transmit, which is worth further investigation. 

The studies of EM exposure transmission design in the literature mainly focus on two aspects, i.e., decreasing the EM exposure with quality of service (QoS) constraints and enhancing the QoS with EM constraints \cite{IEA22,LEA21,JEA22,CCM04,YLH15,YLH17,JYWW22}. For the first aspect, authors in \cite{IEA22} investigated a reconfigurable intelligent surface (RIS) assisted system where the RIS phases and beamforming matrix were jointly designed to reduce the EM exposure. In \cite{LEA21}, tethered unmanned aerial vehicles (TUAVs) were introduced into the network architecture to minimize the EM exposure while ensuring high data rate. Authors in \cite{JEA22} considered the communication with  improper hardware distortion noise and used probabilistic shaping to minimize the EM exposure while achieving the target throughput.

For the second aspect, the work in \cite{YLH15} investigated the precoding design for the capacity with multiple SAR constraints on MIMO uplink. Moreover, the sum-rate analysis with different EM exposure constraints was examined in \cite{YLH17}. The active design of the SAR-constrained SE maximization transmission in metamaterials assisted system was further analyzed in \cite{JYWW22}. The work in \cite{XYN21} investigated the EE maximization transmission strategy for uplink MIMO with SAR constraints. Note that all the previous studies focus on the precoding design for the multiuser scenario without rate split. The EM exposure-aware transmission for the RSMA scheme still requires new algorithm development. In addition, most studies consider the uplink transmission in the criterion of SE maximization and ignore the power cost in the precoding design which may result in low energy efficiency (EE) \cite{YXY20}. Attributed to the economic and ecological concerns, the transmission design for EE maximization is receiving an increasing interest in the literature \cite{XQ13,ZJ15,VON19,YHZ21,ZR22}. Therefore, the EE analysis with EM exposure constraints is also a critical issue for massive MIMO uplink.

Inspired by the aforementioned considerations, we intend to investigate the EM exposure-aware EE maximization design for multiuser massive MIMO uplink RSMA transmissions with statistical CSI. To the authors' best knowledge, this is the first work that proposes to use RSMA for energy and EM exposure efficient communications. The main contributions of this paper are summarized as follows:

\begin{itemize}	
	\item[$\bullet$] We investigate the uplink RSMA transmission design for multiuser massive MIMO systems in the criterion of EE maximization with EM exposure constraints, where each user transmits a superposition of several sub-signals from the split layers. To decouple the transmit covariance matrices and decoding order, we divide the original problem into the equivalent two levels, where the inner one is the optimization of transmit covariance matrices of all layers of users with given decoding permutation and the outer one is to search the optimal decoding order on its feasible set that can maximize the results of the inner optimization. 
	\item[$\bullet$] 
	To address the inner problem, we approximate the ergodic EE by applying the deterministic equivalent (DE) method to reduce the computation complexity. To deal with the non-convex numerators of the objective function, we adopt the minorization-maximization (MM) methods to construct the problem with convex numerators through Taylor expansion linearization. Then, based on the Dinkelbach’s method, we transform the non-convex fractional program into a series of convex sub-problems.
	After that, a modified water-filling algorithm is proposed for the EM aware problem.
	\item[$\bullet$]
	For the outer problem, the optimal decoding order can be obtained using the exhaustive method. To reduce the optimization complexity, we further propose a greedy approach so that the decoding order can be optimized in advance according to channel characteristics. After combining all the above methods, we propose the overall low-complexity EE maximization algorithm for uplink RSMA transmissions with EM exposure constraints. Numerical results demonstrate the effectiveness of our proposed algorithm.
\end{itemize}

The rest of this paper is organized as follows. \secref{sec:system model} describes the model of the uplink RSMA in multiuser massive MIMO systems and formulates the EM exposure-aware EE maximization problem with statistical CSI. \secref{sec:opt_RE_design} presents the overall optimization algorithm for uplink RSMA with EM exposure constraints. In \secref{sec:numerical_results}, numerical results are analyzed for the performance of the proposed algorithm. \secref{sec:conclusion} draws the conclusions of this paper. 

The major notations are list as follows. $\mathbb{E}\{\cdot\}$ denotes the expectation operation, $A\triangleq B$ means the quantity $A$ is defined by $B$. $\mathcal{U}_{\backslash(k,l)}$ represents the set composed of the remaining elements of $\mathcal{U}$ after removing its element $(k,l)$. $\diag{\bQ_{\alpha_k,\beta_{l_k}}}^{K,L_k}_{k,l_k=1}$ denotes the block diagonal matrix where the $\left[\sum\limits_{a=1}^{k-1}L_a+l_k\right]$-th block is $\bQ_{\alpha_k,\beta_{l_k}}$. $\diag{\ba}$ denotes the diagonal matrix composed of elements from vector $\ba$. The operation $(\cdot)!$ and $\odot$ means the factorial and Hadamard product, respectively.  $(x)^+=\max\{x,0\}$. The imaginary unit is represented by $\jmath=\sqrt{-1}$. 

\section{System Model}\label{sec:system model}
\subsection{Rate-Splitting Multiple Access}
\begin{figure}[t]
	\centering
	\includegraphics[width=0.48\textwidth]{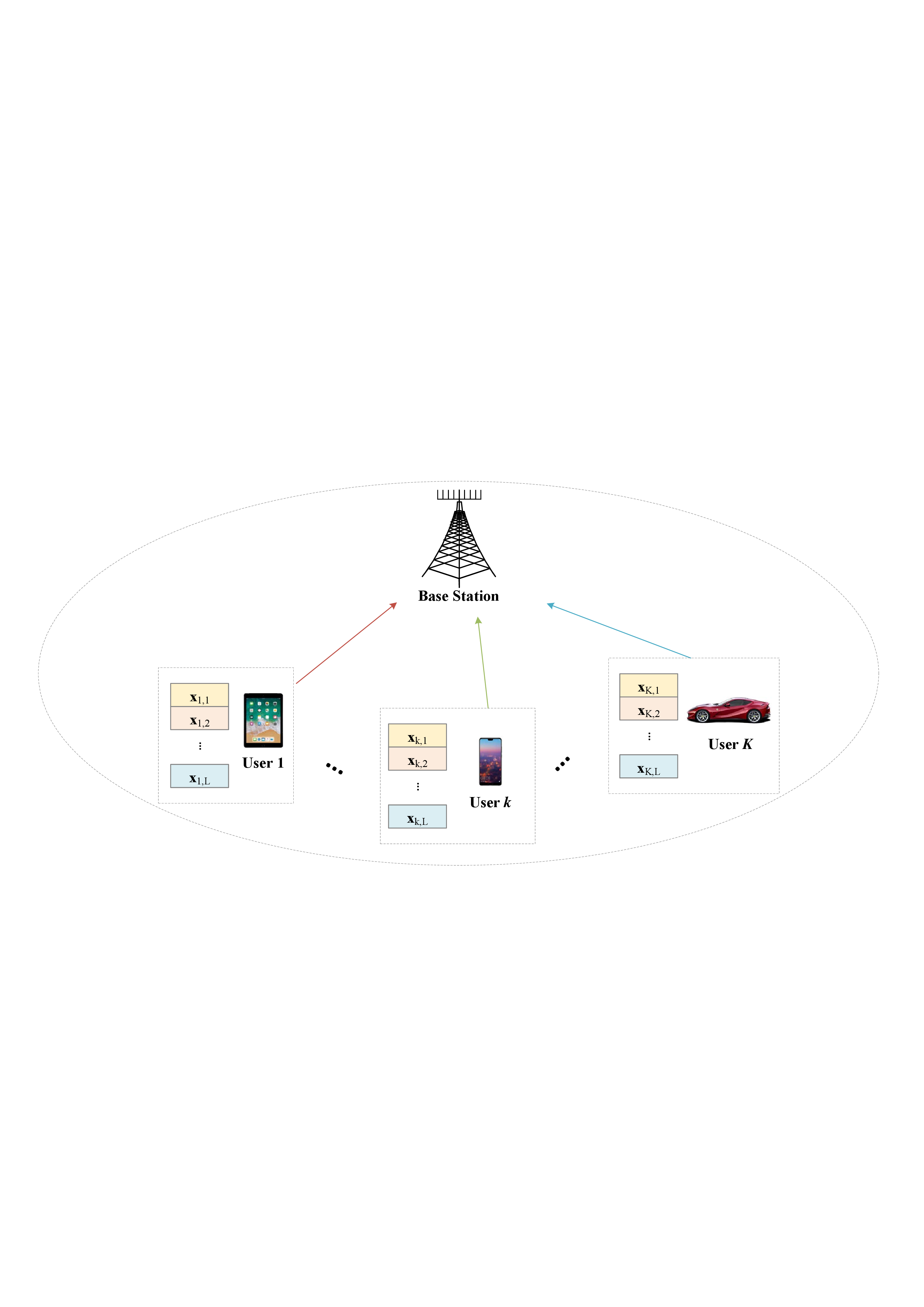}
	\caption{The uplink transmission in a single cell by using rate-splitting multiple access.}
	\label{fig:multi_layer}
\end{figure}
As illustrated in \figref{fig:multi_layer}, we consider a single cell uplink communication of the RSMA system where $K$ users with $N_k$ antennas at each user $k\in\mathcal{K}=\{1,...,K\}$ transmit signals to a $M$-antenna BS simultaneously in the same frequency and time resource. The original data stream $\bs_k\in\C^{N_k\times 1}$ of user $k$ is split into $L$ layers, i.e., $\{\bs_{k,1},...,\bs_{k,L}\}$, where $\bs_{k,l}\in\C^{N_k\times 1}$ is the sub-data stream on layer $l\in\mathcal{L}=\{1,...,L\}$ of user $k$ satisfying $\mathbb{E}\{\bs_{k,l}\}=\mathbf{0}$, $\mathbb{E}\{\bs_{k,l}\bs_{k,l}^H\}=\bI_{N_k}$ and $\mathbb{E}\{\bs_{k,l}\bs_{i,j}^H\}=\mathbf{0}, \forall (k,l)\neq(i,j)$ \cite{ZLN19}. Then, rate-splitting is achieved by precoding these sub-data streams respectively and transmitting the superposition of the precoded signals to the BS \cite{YCS22}. Denote the encoded transmit signal of user $k$ at layer $l$ as
\begin{align}\label{equ:xkl}
\bx_{k,l}=\bF_{k,l}\bs_{k,l}\in\C^{N_k\times 1},
\end{align} 
where $\bF_{k,l}\in\C^{N_k\times N_k}$ is the corresponding transmit precoding matrix. Similarly, $\bx_{k,l}$ is zero mean and independent of other transmit signals, i.e., $\mathbb{E}\{\bx_{k,l}\}=\mathbf{0}$ and $\mathbb{E}\{\bx_{k,l}\bx_{i,j}^H\}=\mathbf{0}, \forall (k,l)\neq(i,j)$, and its covariance matrix is $\bQ_{k,l}=\mathbb{E}\{\bx_{k,l}\bx_{k,l}^H\}$. Then, the transmit signal from user $k$, as the summation of the signals from $L$ layers, is given by 
\begin{align}\label{equ:xk}
\bx_{k}=\sum\limits_{l=1}^L\bx_{k,l}=\sum\limits_{l=1}^L\bF_{k,l}\bs_{k,l}\in\C^{N_k\times 1}.
\end{align}
Correspondingly, the receive signal at the BS is given by 
\begin{align}\label{equ:received_y1}
\by=\sum\limits_{k=1}^{K}\sum\limits_{l=1}^{L}\bH_k\bx_{k,l}+\bn\in \C^{M\times1},
\end{align}  
where $\bH_k\in\C^{M\times N_k}$ is the channel matrix from user $k$ to the BS, $\bn\in\C^{M\times 1}$ denotes the additive noise following circular symmetric complex Gaussian distribution with zero mean and covariance $\sigmatwo\bI_{M}$. Then, the BS decodes $\bx_{k,l},\forall k,l$ from the receive signal $\by$ by using SIC. The decoding order is indicated by an ascending permutation $\bpi=(\pi_{k,l})_{\forall k,l}$, where $\pi_{k,l}\in\mathbb{N}^+$ means the serial number of the decoding order of the sub-signal $\bx_{k,l}$ and satisfies $\pi_{k,l}\neq\pi_{i,j},\forall (k,l)\neq(i,j)$. Specifically, the BS has successfully decoded all the sub-signals ahead of $\pi_{k,l}$ in the decoding permutation and eliminates them in advance, then decodes $\bx_{k,l}$ by treating the remaining sub-signals, except for the desired sub-signal, as interference. Denote $\mathcal{U}=\left\{(k,l)|k\in\mathcal{K},l\in\mathcal{L}\right\}$,
we can write the receive signal at the BS as 
\begin{align}\label{equ:received_y2}
\by=\underbrace{\sum\limits_{(\bar{p},\bar{q})\in\bar{\mathcal{\bm{Q}}}_{k,l}}\bH_{\bar{p}}\bx_{\bar{p},\bar{q}}}_{\text{eliminated in advance}}+\underbrace{\bH_k\bx_{k,l}}_{\text{desired signal}}
+\underbrace{\sum\limits_{(p,q)\in\mathcal{\bm{Q}}_{k,l}}\bH_{p}\bx_{p,q}}_{\text{interference}} +\bn,
\end{align} 
where $\mathcal{\bm{Q}}_{k,l}=\left\{(p,q)\in\mathcal{U}|\pi_{p,q}>\pi_{k,j}\right\}$ and $\bar{\mathcal{\bm{Q}}}_{k,l}=\left\{(\bar{p},\bar{q})\in\mathcal{U}|\pi_{\bar{p},\bar{q}}<\pi_{k,j}\right\}$ is the complementary set of $\mathcal{\bm{Q}}_{k,l}$ in the space of $\mathcal{U}_{\backslash(k,l)}$.

\subsection{System EE}
Consider that transmitters only have the statistical CSI for uplink channels, whose spatial correlations are described by the jointly correlated Rayleigh fading model. Then, $\bH_k$ can be characterized as \cite{GJL09}
\begin{align}\label{equ:Hk}
\bH_k=\bU_k\widetilde{\bH}_k\bV_k^H\in\C^{M\times N_k},
\end{align}
%and termed the uplink channel of beam domain,
where $\widetilde{\bH}_k\in\C^{M\times N_k}$ is the beam domain channel matrix with zero-mean and independently distributed elements, $\bU_k\in\C^{M\times M}$ and $\bV_k\in\C^{N_k\times N_k}$ are deterministic unitary matrices. Then, the statistical CSI of $\bH_k$ can be described by the coupling matrix between the transmit and receive eigen-directions, which is defined as \cite{WJW11}
\begin{align}\label{equ:Omegak}
\Omegak=\mathbb{E}\left\{\widetilde{\bH}_k\odot\widetilde{\bH}_k^*\right\}\in\C^{M\times N_k}.
\end{align}
In particular, when $M\to\infty$ in our considered massive MIMO systems, $\bU_k$ in \eqref{equ:Hk} becomes asymptotically identical for all users, which can be expressed as \cite{ANA13,YGX15}
\begin{align}\label{equ:U}
\bU_{k}\overset{M\to\infty}{=}\bU\in\C^{M\times M}, \quad \forall k,
\end{align}
%,CYL21
where $\bU$ is irrelevant to the locations of users and only depends on the topologies of BS antenna array \cite{ANA13}. For example, when the BS employs the uniform linear array (ULA) antenna spacing of half-wavelength, the discrete Fourier transform (DFT) matrix takes a good approximation of $\bU$ \cite{YGX15,YHZ21}.
According to \eqref{equ:received_y2}, the BS treat $\bn_{k,l}=\sum_{(p,q)\in\mathcal{\bm{Q}}_{k,l}}\bH_{p}\bx_{p,q}+\bn$ as the aggregate interference-plus-noise that follows Gaussian distribution for a worst-case design when decoding the desired signal $\bx_{k,l}$ \cite{HH03}. Then, the achievable ergodic rate of the sub-signal $\bx_{k,l}$ can be formulated by
\begin{align}\label{equ:Rklori}
&R_{k,l}^{\mathrm{ach}}=\mathbb{E}\left\{\log\det\left(\bI_{M}+\bH_k\bQ_{k,l}\bH_k^H\bC_{k,l}^{-1}\right)\right\}\ntb
&=\mathbb{E}\left\{\log\det\left(\bC_{k,l}+\bH_k\bQ_{k,l}\bH_k^H\right)\right\}-\mathbb{E}\left\{\log\det\left(\bC_{k,l}\right)\right\},
\end{align}
where $\bC_{k,l}$ denotes the covariance matrix of the interference-plus-noise $\bn_{k,l}$ expressed by
\begin{align}\label{equ:IPN1}
\bC_{k,l}=\sigmatwo\bI_{M}+\sum\limits_{(p,q)\in\mathcal{\bm{Q}}_{k,l}}\bH_p\bQ_{p,q}\bH_p^H\in\C^{M\times M}.
\end{align}
Due to the channel hardening effect in the transmission of massive MIMO, we can approximate the ergodic rate of the decoded signal $\bx_{k,l}$ in \eqref{equ:Rklori} by \cite{YXZ20,LGZ19,WGW18}
\begin{align}\label{equ:Rkl}
R_{k,l}&=\mathbb{E}\left\{\log\det\left(\wbK_{k,l}+\bH_k\bQ_{k,l}\bH_k^H\right)\right\}-\log\det\left(\wbK_{k,l}\right),
\end{align}
where 
\begin{align}\label{equ:wbKkl}
\wbK_{k,l}=\mathbb{E}\{\bC_{k,l}\}=&\sum\limits_{(p,q)\in\mathcal{\bm{Q}}_{k,l}}\bU\mathbb{E}\{\wbH_p\bV_p^H\bQ_{p,q}\bV_p\wbH_p^H\}\bU^H\ntb
&+\sigmatwo\bI_{M}.
\end{align}
Denote the matrix-valued function on $\bX\in\C^{N_k\times N_k}$ as 
\begin{align}\label{equ:wbTheta}
\wbTheta_k(\bX)\triangleq\mathbb{E}\left\{\widetilde{\bH}_k\bV_k^H\bX\bV_k\widetilde{\bH}_k^H\right\}\in\C^{M\times M}. 
\end{align}
Based on the statistical characteristics of the channel matrix in the beam domain, it can be derived that $\wbTheta(\bX)$ is a diagonal matrix, where the diagonal elements are obtained by
\begin{align}\label{equ:wbTheta_ele}
\left[\widetilde{\bTheta}_k(\bX)\right]_{ii}&=\sum\limits_{j=1}^{N_k}\left[\Omegak\right]_{ij}\left[\bV_k^H\bX\bV_k\right]_{jj}, \forall i=1,...,M. 
\end{align}
Note that by rewriting $\bH_k$ according to \eqref{equ:Hk} and exploiting the Sylvester’s determinant identity, i.e., $\det(\bI+\bA\bB)=\det(\bI+\bB\bA)$, the approximation rate in \eqref{equ:Rkl} is equivalent to
\begin{align}\label{equ:reRkl}
R_{k,l}=&\mathbb{E}\left\{\log\det\left(\bK_{k,l}+\widetilde{\bH}_k\bV_k^H\bQ_{k,l}\bV_k\widetilde{\bH}_k^H\right)\right\}\ntb
&-\log\det\left(\bK_{k,l}\right),
\end{align}
where $\bK_{k,l}$ is given by
\begin{align}\label{equ:Kkl}
\bK_{k,l}=\bU^H\wbK_{k,l}\bU=\sigmatwo\bI_{M}+\sum\limits_{(p,q)\in\mathcal{\bm{Q}}_{k,l}}\widetilde{\bTheta}_p(\bQ_{p,q}).
\end{align}

In RSMA systems, the total power consumption is described by an affine model comprised of three parts \cite{MMD22,XQ13}, i.e., 
\begin{align}\label{equ:PQ}
P(\bQ)=\sum\limits_{k=1}^K\left(\xi_{k}\sum\limits_{l=1}^L\tr{\bQ_{k,l}}+P_{c,k}\right)+P_{\mathrm{BS}},
\end{align}
where $\bQ\triangleq\diag{\bQ_{k,l}}_{K,L}$ represents the aggregate covariance matrix for all layers and users, $\xi_{k}(>1)$ denotes the inverse of the power amplifier inefficiency at user $k$, $P_{c,k}$ is the dynamic power dissipation of user $k$, and $P_{\mathrm{BS}}$ incorporates the circuit power consumption at the BS \cite{WJW11}. Denote the communication bandwidth as $W$, then the EE of the system can be written as
\begin{align}\label{equ:EE}
\mathrm{EE}=W\frac{\sum_{(k,l)\in\mathcal{U}}R_{k,l}}{P(\bQ)}.
\end{align}

\subsection{EM Exposure Model}
In practical uplink communications, the transmit signals are not only power constrained but also restricted by specific EM exposure level \cite{YLH17}. In general, the power constraints are described by $\sum\limits_{l=1}^{L}\tr{\bQ_{k,l}}\leq\Pmaxk$ for all user $k$, where $\Pmaxk$ represents the power budget of user $k$. Moreover, the EM exposure at transmitters is usually measured by SAR, which can be modeled as a quantity averaged over the transmit signals with a time-averaged quadratic constraint given by \cite{YLH15}
%Moreover, the EM exposure at transmitters is usually measured by SAR, whose main difference from the power exits that SAR is a function of the phase difference between any two transmit antennas, while power is only related to the modulus of the transmit signal \cite{HLY13}. The SAR model is based on experimental studies, which can be characterized by a sinusoidal function of the phase difference \cite{HL12,CCM04}. By introducing the SAR matrix $\bR_{k,a}$, the SAR can be modeled as a quantity averaged over the transmit signals with a time-averaged quadratic constraint given by \cite{YLH15}
\begin{align}\label{equ:SAR_ka}
\mathrm{SAR}_{k,a}&=\sum\limits_{l=1}^{L}\mathbb{E}\{\tr{\bx_{k,l}^H\Rka\bx_{k,l}}\}=\sum\limits_{l=1}^{L}\tr{\Rka\bQ_{k,l}}\ntb
&\leq D_{k,a},\ a=1,2,...,A_k,
\end{align}
where the subscripts $k,a$ represent the $k$-th user and $a$-th partial body of mathematical quantities, $\bR_{k,a}$ is the SAR matrix, which fully describes the dependence of SAR measurements on transmit signals with the unit of each entry as $\mathrm{kg}^{-1}$ \cite{YLH17}, $D_{k,a}$ denotes the SAR budget, and $A_k$ means the number of the partial body exposed to the EM radiation. According to \cite{CCM04}, with given transmit antennas, the SAR measurements vary slightly over a fairly wide range of frequencies (1.8 – 2 GHz). For the uplink communication where $W<200$ MHz, the SAR measurements in different frequencies are almost the same, which allows all the carriers in our considered RSMA system to share the same SAR matrix \cite{CCM04}. Note that $\bR_{k,a}$ is relatively stable for a device within the given model \cite{HLY13}. In practice, SAR matrices can be reported by users to the BS after the transmission link is established.

\subsection{Problem Formulation}
In this paper, we investigate the EM exposure-aware transmission strategy design for RSMA uplink of multiuser massive MIMO under the EE maximization criterion, where we optimize the transmit covariance matrices of sub-signals $\{\bQ_{k,l}\}_{\forall k,l}$ at the users and the decoding order $\bpi$ at the BS to maximize the system EE, i.e.,
\begin{subequations}\label{eq:problem1}
	\begin{align}
	\cP_1:\quad&\underset{\{\bQ_{k,l}\}_{\forall k,l},\bpi} \max \quad \mathrm{EE}(\bQ,\bpi), \label{p1a}\\
	&{\mathrm{s.t.}} \ \sum\limits_{l=1}^{L}\tr{\bQ_{k,l}}\leq\Pmaxk,\quad \bQ_{k,l} \succeq \mathbf{0}, \label{p1b}\\
	&\quad\ \ \sum\limits_{l=1}^{L}\tr{\bR_{k,a}\bQ_{k,l}}\leq D_{k,a}, \quad\forall k,a, \label{p1c}\\
	&\quad\ \ \bpi\in\bPi, \label{p1d}
	\end{align}
\end{subequations}
where $\bPi$ is the set composed of all possible decoding order of sub-signals and contains a total of $(KL)!$ discrete elements. 

Note that the challenges in tackling $\cP_1$ come from two aspects. The first one lies in the objective function, which is non-convex on both covariance matrices and decoding order. In addition, because of the expectation operations of \eqref{equ:reRkl}, the computational cost becomes enormous if the Monte Carlo method is adopted to calculate the average EE in the optimization. The second one lies in the constraints. In particular. the introduction of SAR constraints, i.e., \eqref{p1c}, brings great challenges in obtaining the optimal covariance matrices, where we can not determine the optimal transmit direction and then transform $\cP_1$ into the conventional power allocation problem. Moreover, it can be observed that the discrete variable $\bpi$, which makes $\cP_1$ a non-convex mixed integer problem, further complicates the problem. In the following, we will develop algorithms to address problem \eqref{eq:problem1}.

\section{EM Exposure-Aware EE Optimization}\label{sec:opt_RE_design}
In this section, we investigate the strategy design of problem $\cP_1$ under the criterion of EE maximization. Note that it is challenging to figure out the optimal solution of $\cP_1$ beacuse of the decoding order constraint \eqref{p1d}. To tackle this problem, we rewrite \eqref{p1a} as
\begin{align}\label{equ:rewirte}
\underset{\bpi}\max \quad \underset{\{\bQ_{k,l}\}_{\forall k,l}}\max \quad \mathrm{EE}(\bQ,\bpi).
\end{align}
By fixing the permutation $\bpi$, the inner optimization of \eqref{equ:rewirte} can be expressed as
\begin{subequations}\label{eq:problemlow}
	\begin{align}
	\cP_2^{\mathrm{in}}:\quad &f(\bpi)=\underset{\{\bQ_{k,l}\}_{\forall k,l}}\max \quad \mathrm{EE}(\bQ,\bpi), \label{plowa}\\
	&{\mathrm{s.t.}} \ \sum\limits_{l=1}^{L}\tr{\bQ_{k,l}}\leq\Pmaxk,\quad \bQ_{k,l} \succeq \mathbf{0}, \label{plowb}\\
	&\quad\ \  \sum\limits_{l=1}^{L}\tr{\bR_{k,a}\bQ_{k,l}}\leq D_{k,a}, \quad\forall k,a. \label{plowc}
	\end{align}
\end{subequations}
Then, setting the maximal value of the inner optimization to $f(\bpi)$, $\cP_1$ can be formulated as the equivalent problem given by
\begin{subequations}\label{eq:problemup}
	\begin{align}
	\cP_2^{\mathrm{out}}:\quad &\underset{\bpi}\max \quad f(\bpi), \label{pupa}\\
	&{\mathrm{s.t.}} \quad \bpi\in\bPi.  \label{pupb}
	\end{align}
\end{subequations}
The optimal decoding order can be obtained by exhaustive search, which will introduce high complexity to the overall optimization \cite{YCS22}. With this in mind, we intend to figure out a greedy approach where the decoding permutation can be determined in advance. 
%i.e., traversing through all the elements of $\bPi$ to search for the optimal $\bpi$ that maximizes the system EE \cite{YCS22}. 
In the following, we first investigate the optimization of transmit covariance matrices with a given decoding order. 

\subsection{Deterministic Equivalence}
Assuming that the decoding order $\bpi$ is fixed within its feasible set, we can determine $\bK_{k,l},\forall k,l$ by \eqref{equ:Kkl} and rewrite the objective function of \eqref{eq:problemlow} as 
\begin{align}\label{equ:etaEE}
\eta_{\mathrm{EE}}(\bQ)=\frac{\sum_{(k,l)\in\mathcal{U}}r_{k,l}^+(\bQ)-r_{k,l}^-(\bQ)}{P(\bQ)},
\end{align}
where
\begin{align}
r_{k,l}^+(\bQ)&=\mathbb{E}\left\{\log\det\left(\bK_{k,l}+\widetilde{\bH}_k\bV_k^H\bQ_{k,l}\bV_k\widetilde{\bH}_k^H\right)\right\}, \label{equ:rkl+}\\
r_{k,l}^-(\bQ)&=\log\det\left(\sigmatwo\bI_{M}+\sum\limits_{(p,q)\in\mathcal{\bm{Q}}_{k,l}}\widetilde{\bTheta}_p(\bQ_{p,q})\right). \label{equ:rkl-}
\end{align}

It is noteworthy that problem $\cP_2^{\mathrm{in}}$ with the optimization function of \eqref{equ:etaEE} involves random elements in the calculation. However, using stochastic programming approaches like the Monte-Carlo method will incur huge computation complexity \cite{CD11}. To mitigate the computational burden, we approximate the ergodic EE by applying the DE method, which is a low complexity approach for calculating the expected values without averaging \cite{LGX16}. According to the large-dimensional random matrix theory, the DE provides the deterministic approximation of the function with random matrices, which is asymptotically accurate when the dimensions of matrices increase to infinity at a fixed rate \cite{WJW11,LGX16}. The DE method is achieved by introducing several auxiliary variables and iteratively calculating the objective function. For further simplified calculation, we denote 
\begin{align}\label{equ:bTheta}
\bTheta_k(\bX)\triangleq\mathbb{E}\left\{\widetilde{\bH}_k^H\bX\widetilde{\bH}_k\right\}\in\C^{N_k\times N_k}.
\end{align}
Similarly, utilizing the property that the elements of $\widetilde{\bH}_k$ are zero-mean and independently distributed, $\bTheta_k(\bX)$ is a $N_k\times N_k$-dimensional diagonal matrix with the elements
\begin{align}\label{equ:bTheta_ele}
\left[\bTheta_k(\bX)\right]_{jj}&=\sum\limits_{i=1}^{M}\left[\Omegak\right]_{ij}\left[\bX\right]_{ii}, \forall j=1,...,N_k, 
\end{align}
where $\Omegak$ is defined in \eqref{equ:Omegak}. Then, \eqref{equ:rkl+} can be well approximated at a low computational level by using the eigenmode coupling matrix $\Omegak$ as follows \cite{LGX16}
\begin{align}\label{equ:DERkl}
r_{k,l}^+(\bQ)&\approx R_{k,l}^+(\bQ)=\log\det\left(\bI_{N_k}+\bGamma_{k,l}\bQ_{k,l}\right)\ntb
&+\log\det\left(\wbGamma_{k,l}+\bK_{k,l}\right)-\tr{\bI_{M}-\wbPhi_{k,l}^{-1}},
\end{align}
where $\bGamma_{k,l}$, $\wbGamma_{k,l}$ and $\wbPhi_{k,l}$ are DE parameters respectively given by
\begin{subequations}\label{eq:DE}
	\begin{align}
	&\bGamma_{k,l}=\bV_k\bTheta_k(\wbPhi_{k,l}^{-1}\bK_{k,l}^{-1})\bV_k^H,\label{equ:DEpara1}\\
	&\wbGamma_{k,l}=\wbTheta_k(\bQ_{k,l}^{\frac{1}{2}}\bPhi_{k,l}^{-1}\bQ_{k,l}^{\frac{1}{2}}),\label{equ:DEpara2}\\
	&\wbPhi_{k,l}=\bI_{M}+\wbGamma_{k,l}\bK_{k,l}^{-1},\label{equ:DEpara3}\\
	&\bPhi_{k,l}=\bI_{N_k}+\bQ_{k,l}^{\frac{1}{2}}\bGamma_{k,l}\bQ_{k,l}^{\frac{1}{2}}.\label{equ:DEpara4}
	\end{align}
\end{subequations}
Substituting \eqref{equ:DERkl} into \eqref{equ:reRkl}, the DE expression of the ergodic rate for user $k$ at layer $l$ can be expressed as
\begin{align}\label{equ:RKlbar}
\overline{R}_{k,l}(\bQ)=R_{k,l}^+(\bQ)-r_{k,l}^-(\bQ).
\end{align}

As elucidated above, the DE equation $R_{k,l}^+(\bQ)$ is determined by DE parameters, which are related to the aggregate transmit covariance matrix $\bQ$ and require iterative calculations. With an initial value of $\bPhi_{k,l}$, the DE parameters $\bGamma_{k,l}$, $\wbGamma_{k,l}$ and $\wbPhi_{k,l}$ that are used to approximate $r_{k,l}^+(\bQ)$ can be obtained by cyclically updating $\bPhi_{k,l}$ and $\wbPhi_{k,l}$ through \eqref{equ:DEpara1}--\eqref{equ:DEpara4}. Therefore, an iterative algorithm for calculating the DE of the ergodic SE is proposed in \textbf{\alref{alg:DEs}}.
\begin{algorithm}[h]
	\caption{Computation of DE Parameters of $r_{k,l}^{+}(\bQ)$}
	\label{alg:DEs}
	\begin{algorithmic}[1]
		\Require Initial $\left\{\bPhi_{k,l}^{(0)}\right\}_{\forall k,l}$, channel statistics $\left\{\bOmega_{k}\right\}_{\forall k}$, feasible covariance matrix $\bQ$ and permutation $\bpi$, iteration threshold $\epsilon_1$.
		\Ensure The DE parameters $\bGamma_{k,l}$, $\wbGamma_{k,l}$ and $\wbPhi_{k,l}$, $\forall (k,l)\in\mathcal{U}$.
		\State Initialize $\left\{\bPhi_{k,l}^{(0)}\right\}_{\forall k,l}$, iteration index $u=0$, threshold $\epsilon_1$.
		\For{all $(k,l)$ : $\mathcal{U}$}
		\State Reset iteration index $u = 0$.
		\Repeat
		\State Calculate $\wbPhi_{k,l}^{(u)}$ by \eqref{equ:DEpara2} and \eqref{equ:DEpara3} with given $\bPhi_{k,l}^{(u)}$.
		\State Update $\bPhi_{k,l}^{(u+1)}$ by \eqref{equ:DEpara1} and \eqref{equ:DEpara4} with given $\wbPhi_{k,l}^{(u)}$.
		\State Set $u=u+1$.
		\Until $\left|\left|\bPhi_{k,l}^{(u)}-\bPhi_{k,l}^{(u-1)}\right|\right|^2_{\mathrm{F}}\leq \epsilon_1$.
		\State Return $\bPhi_{k,l}=\bPhi_{k,l}^{(u)}$ and $\wbPhi_{k,l}=\wbPhi_{k,l}^{(u)}$.
		\State Calculate $\bGamma_{k,l}$ and $\wbGamma_{k,l}$ by \eqref{equ:DEpara1} and \eqref{equ:DEpara2}, respectively with given $\bPhi_{k,l}^{(u)}$ and $\wbPhi_{k,l}^{(u)}$.
		\EndFor
		\State Return $\left\{\bGamma_{k,l}\right\}_{\forall k,l}$, $\left\{\wbGamma_{k,l}\right\}_{\forall k,l}$, and $\left\{\wbPhi_{k,l}\right\}_{\forall k,l}$.
	\end{algorithmic}
\end{algorithm}

Note that the DE method reduces the computational complexity compared with the Monte-Carlo method, which requires a great number of repeated optimization problems to average the solutions. In addition, due to the fact $M\gg N_k,\forall k$ for common RSMA uplink massive MIMO systems, the calculation of DE parameters is of low complexity despite the operation of matrix inversion. It can be observed that $\bK_{k,l}$, $\wbGamma_{k,l}$ and $\wbPhi_{k,l}$ are diagonal matrices so that their inverses can be easily obtained, whose complexity is negligible for the overall optimization process.

Then, by approximating $\etaSE(\bQ)$ with its DE expression, which is defined by $\etaEEb(\bQ)$, the asymptotic EE maximization problem without expectation operation can be expressed as 
\begin{subequations}\label{eq:problem3}
	\begin{align}
	\cP_3:\ \underset{\{\bQ_{k,l}\}} \max &\  	\etaEEb(\bQ)=\frac{\sum_{(k,l)\in\mathcal{U}}\overline{R}_{k,l}(\bQ)}{P(\bQ)}, \label{p3a}\\
	{\mathrm{s.t.}} &\ \sum\limits_{l=1}^{L}\tr{\bQ_{k,l}}\leq\Pmaxk,\quad \bQ_{k,l} \succeq \mathbf{0}, \label{p3b}\\
	&\ \sum\limits_{l=1}^{L}\tr{\bR_{k,a}\bQ_{k,l}}\leq D_{k,a}, \quad\forall k,a. \label{p3c}
	\end{align}
\end{subequations} 

\subsection{Low Complexity Precoding Design for RSMA}\label{sbsec:MMDinkelWaterfill}
Generally, the fractional programming approaches are adopted to handle the problem $\cP_3$ that contains one fractional objective function. However, $R_{k,l}^+(\bQ)$ and $r_{k,l}^-(\bQ)$ are both concave over $\bQ$ as indicated in \cite{ZJ15,BV04}, which results in the non-convexity of the numerator in \eqref{p3a}. Then, $\cP_3$ is an NP-hard problem and can not be guaranteed the global optimal solution within a polynomial-time complexity \cite{ZJ15}. Therefore, directly using the classical fractional programming methods for solving the EE maximization problem will exponentially increase the complexity, which calls for an efficient algorithm to tackle $\cP_3$. To this end, we resort to the MM method to handle $\cP_3$ by solving a series of subproblems that successively approximate the original problem.

As elaborated in \cite{MW78,SBP17}, the MM procedure is an iterative optimization procedure used to asymptotically solve the non-convex program, where the main idea is to construct a series of solvable lower bound functions with the utilize of local maximizer points of each iteration. Note that the non-convexity of \eqref{p3a} comes from the subtraction of two convex functions, which inspires us to construct the convex function by linearizing the second term. To this end, we denote the derivative of $r_{k,l}^-(\bQ)$ over each $\bQ_{p,q}$ for any $(p,q)\in\mathcal{\bm{Q}}_{k,l}$ as $\bDelta_{p,q}^{k,l}\triangleq\frac{\partial r_{k,l}^{-}(\bQ)}{\partial\bQ_{p,q}}$. It is shown as a conjugate symmetric matrix and can be formulated as
\begin{align}\label{equ:partialrkl-}
\bDelta_{p,q}^{k,l}&=\bV_p\mathbb{E}\{\widetilde{\bH}_p^H\bK_{k,l}^{-1}\widetilde{\bH}_p\}\bV_p^H\ntb
&=\sum\limits_{i=1}^{M}\frac{\bV_p\bW_{p,i}\bV_p^H}{\sigmatwo+\sum\limits_{(p,q)\in\mathcal{\bm{Q}}_{k,l}}\tr{\bV_p\bW_{p,i}\bV_p^H\bQ_{p,q}}},
\end{align}
where
\begin{align}\label{equ:Wkl}
\bW_{p,i}\triangleq\diag{\left[\bOmega_p\right]_{i1},...,\left[\bOmega_p\right]_{ij}}_{j=1}^{N_k}.
\end{align}
Then, the approximate solution of $\cP_3$ can be derived by applying the iterative MM method, where the local maximizer of the $\ell$-th subproblem can be expressed as
\begin{subequations}\label{eq:problem4}
	\begin{align}
	\cP_4^{(\ell)}:\bQ^{(\ell+1)}&=\argmax{\{\bQ_{k,l}\}_{\forall k,l}} \ \frac{\sum_{(k,l)\in\mathcal{U}}\widetilde{R}_{k,l}(\bQ|\bQ^{(\ell)})}{P(\bQ)},\label{eq:p4aMM}\\
	{\mathrm{s.t.}} &\ \sum\limits_{l=1}^{L}\tr{\bQ_{k,l}}\leq\Pmaxk,\quad \bQ_{k,l} \succeq \mathbf{0}, \label{p4b}\\
	&\ \sum\limits_{l=1}^{L}\tr{\bR_{k,a}\bQ_{k,l}}\leq D_{k,a}, \quad\forall k,a. \label{p4c}
	\end{align}
\end{subequations} 
In $\cP_4^{(\ell)}$, $\widetilde{R}_{k,l}(\bQ|\bQ^{(\ell)})$ is the first-order Taylor expansion of $\overline{R}_{k,l}(\bQ)$ on the point $\bQ^{(\ell)}$ given by
\begin{align}\label{equ:tildeRKl}
\widetilde{R}_{k,l}(\bQ|\bQ^{(\ell)})&=R_{k,l}^+(\bQ)-r_{k,l}^-(\bQ^{(\ell)})-\ntb
&\sum\limits_{(p,q)\in\mathcal{\bm{Q}}_{k,l}}\tr{(\bDelta_{p,q}^{k,l})^{(\ell)}(\bQ_{p,q}-\bQ_{p,q}^{(\ell)})}, 
\end{align}
where $\ell\in\mathbb{N}$ denotes the number of iterations and 
$(\bDelta_{p,q}^{k,l})^{(\ell)}$ is the value of $\bDelta_{p,q}^{k,l}$ at point $\bQ_{p,q}^{(\ell)}$. According to \cite{MW78}, the surrogate function $\widetilde{R}_{k,l}(\bQ|\bQ^{(\ell)})$ constructed by the optimization results of the previous iteration complies with the MM iteration condition, which guarantees that $\left\{\bQ^{(\ell)}\right\}_{\ell=0}^{\infty}$ will finally converge to the suboptimal solution of $\cP_3$.

Consider that the numerator and denominator of \eqref{eq:p4aMM} are concave and convex over $\bQ$, respectively. In this paper, we adopt Dinkelbach’s method to transform the concave-convex fractional programming into a sequence of convex problems. Specifically, each $\cP_4^{(\ell)}$ can be equivalently addressed by iteratively handling the maximization subproblems as \cite{ZJ15}
\begin{subequations}\label{eq:problem5}
	\begin{align}
	\cP_5^{(\ell),[t]}:\bQ^{(\ell),[t+1]}&=\argmax{\{\bQ_{k,l}\}_{\forall k,l}} \ \sum\limits_{(k,l)\in\mathcal{U}}\widetilde{R}_{k,l}(\bQ|\bQ^{(\ell)})\ntb
	&\qquad\qquad\qquad\qquad-\eta^{(\ell),[t]}P(\bQ),\label{p5a}\\
	{\mathrm{s.t.}} \ &\sum\limits_{l=1}^{L}\tr{\bQ_{k,l}}\leq\Pmaxk,\  \bQ_{k,l} \succeq \mathbf{0}, \label{p5b}\\
	\ &\sum\limits_{l=1}^{L}\tr{\bR_{k,a}\bQ_{k,l}}\leq D_{k,a}, \ \ \forall k,a, \label{p5c}
	\end{align}
\end{subequations} 
where $t$ denotes the iteration index and $\eta^{(\ell),[t]}$ is updated by
\begin{align}\label{equ:eta}
\eta^{(\ell),[t]}=\frac{\sum_{(k,l)\in\mathcal{U}}\widetilde{R}_{k,l}(\bQ^{(\ell),[t]}|\bQ^{(\ell)})}{P(\bQ^{(\ell),[t]})}.
\end{align}
It can be proved that Dinkelbach's transform finally converges to the globally optimal solution of $\cP_4^{(\ell)}$ \cite{ZJ15,SY18}. With the hybrid use of the MM procedure and Dinkelbach's transform, the approximate solution of problem \eqref{eq:problemlow} containing the nonconvex objective function can be obtained by dealing with a series of convex problems formulated by $\cP_5^{(\ell),[t]}$, which reduces the problem-solving complexity significantly. Note that problem \eqref{eq:problem5} can be solved using either convex optimization tools \cite{BV04,ZJ15} or the modified water-filling scheme proposed later in \secref{sbsec:MMDinkelWaterfill}. 
Combining the above MM and Dinkelbach's methods, the EE maximization precoding design for uplink RSMA is detailed in \textbf{\alref{alg:MMDinkel}}.
\begin{algorithm}[h]
	\caption{EE Maximum Iterative Algorithm for Uplink RSMA}
	\label{alg:MMDinkel}
	\begin{algorithmic}[1]
		\Require Initial covariance matrices $\left\{\bQ_{k,l}^{(0)}\right\}_{\forall k,l}$, channel statistics $\left\{\bOmega_{k}\right\}_{\forall k}$, feasible permutation $\bpi$, iteration thresholds $\epsilon_2$, $\epsilon_3$.
		\Ensure The approximate solution $\left\{\bQ_{k,l}^{\bpi}\right\}_{\forall k,l}$ of $\cP_3$. 
		\State Initialize the iteration index $\ell=0$, $\etaEEb^{(\ell)}=0$.
		\Repeat 
		%		\State Obtain the DE parameters by \textbf{\alref{alg:DEs}}.
		\State Obtain $(\bDelta_{p,q}^{k,l})^{(\ell)}$, $\forall(p,q)\in\mathcal{\bm{Q}}_{k,l}$ by \eqref{equ:partialrkl-} and \eqref{equ:Wkl}.
		\State Set the interation index $t=0$.
		\State Initialize $\left\{\bQ_{k,l}^{(\ell),[t]}\right\}_{\forall k,l}=\left\{\bQ_{k,l}^{(\ell)}\right\}_{\forall k,l}$ and calculate the auxiliary value $\eta^{(\ell),[t]}$ by \eqref{equ:DERkl}, \eqref{equ:tildeRKl} and \eqref{equ:eta}.
		\Repeat
		\State Obtain $\bQ^{(\ell),[t+1]}$ by solving the problem \eqref{eq:problem5}.
		\State Set $t=t+1$.
		\State Update $\eta^{(\ell),[t]}$ by \eqref{equ:eta}.
		\Until{$\left|\eta^{(\ell),[t]}-\eta^{(\ell),[t-1]}\right|<\epsilon_2$.}
		\State Return $\left\{\bQ_{k,l}^{(\ell+1)}\right\}_{\forall k,l}=\left\{\bQ_{k,l}^{(\ell),[t]}\right\}_{\forall k,l}$.
		\State Set $\ell=\ell+1$.
		\State Update the objective function of the $\ell$-th subproblem in the MM procedure by
		\begin{align}
		\etaEEb^{(\ell)}=\frac{\sum_{(k,l)\in\mathcal{U}}\widetilde{R}_{k,l}(\bQ^{(\ell)}|\bQ^{(\ell)})}{P(\bQ^{(\ell)})}.
		\end{align}
		\Until{$\left|\etaEEb^{(\ell)}-\etaEEb^{(\ell-1)}\right|<\epsilon_3$.}
		\State Return $\left\{\bQ_{k,l}^{\bpi}\right\}_{\forall k,l}=\left\{\bQ_{k,l}^{(\ell)}\right\}_{\forall k,l}$.
	\end{algorithmic}
\end{algorithm}

By introducing Lagrange multipliers $\{\mu_k\}_{\forall k}$ and $\{\lambda_{k,a}\}_{\forall k,a}$ corresponding to the dual variables of constraints \eqref{p5b} and \eqref{p5c}, respectively, we derive the dual function of problem \eqref{eq:problem5} as
\begin{align}\label{eq:Lagrange}
&\mathcal{LA}(\bQ,\{\mu_k\},\{\lambda_{k,a}\})=\sum\limits_{(k,l)\in\mathcal{U}}\widetilde{R}_{k,l}(\bQ|\bQ^{(\ell)})-\eta^{(\ell),[t]}P(\bQ)\ntb
&\qquad\qquad-\sum\limits_{(k,l)\in\mathcal{U}}\mu_k\left(\tr{\bQ_{k,l}}-\Pmaxk\right)\ntb
&\qquad\qquad-\sum\limits_{(k,l)\in\mathcal{U}}\sum\limits_{a=1}^A\lambda_{k,a}\left(\tr{\bR_{k,a}\bQ_{k,l}}-D_{k,a}\right),
%+\sum\limits_{k=1}^{K}\sum\limits_{l=1}^{L}\tr{\bPsi_{k,l}\bQ_{k,l}},
\end{align}
where $\mu_k\geq0$ and $\lambda_{k,a}\geq0$, $\forall k,a$. Then, the Lagrange dual problem of $\cP_5^{(\ell),[t]}$ can be expressed as 
\begin{align}\label{equ:dual_problem}
\underset{\mu_k\geq 0,\forall k,\atop\lambda_{k,a}\geq 0,\forall k,a}\min\ \underset{\bQ\succeq\mathbf{0}}\max \quad \mathcal{LA}(\bQ,\{\mu_k\},\{\lambda_{k,a}\}).
\end{align}

\begin{prop}\label{Prop:Water-Filling}
	Constructing the optimal auxiliary matrix 
	\begin{align}\label{equ:Skl}
	\bS_{k,l}^{\star}=&\sum\limits_{(\bar{p},\bar{q})\in\bar{\mathcal{\bm{Q}}}_{k,l}}(\bDelta_{k,l}^{\bar{p},\bar{q}})^{(\ell)}+(\eta^{(\ell),[t]}\xi_{k}+\mu_k^{\star})\bI_{N_k}+\ntb
	&\sum\limits_{a=1}^A\lambda_{k,a}^{\star}\bR_{k,a},
	\end{align}	
	where $\{\mu_k^{\star}\}_{\forall k}$ and $\{\lambda_{k,a}^{\star}\}_{\forall k,a}$ are optimal dual variables of	power and SAR constraints in problem \eqref{eq:problem5}, respectively. Then, the optimal solutions of $\cP_5^{(\ell),[t]}$ can be derived by a modified water-filling scheme over $(\bS_{k,l}^{\star})^{-\frac{1}{2}}\bGamma_{k,l}(\bS_{k,l}^{\star})^{-\frac{1}{2}}$ with the eigenvalue decomposition
	\begin{align}\label{equ:SGammaDecom}
	(\bS_{k,l}^{\star})^{-\frac{1}{2}}\bGamma_{k,l}(\bS_{k,l}^{\star})^{-\frac{1}{2}}=\widetilde{\bU}_{k,l}\bSigma_{k,l}\widetilde{\bU}_{k,l}^H,
	\end{align}
	where $\widetilde{\bU}_{k,l}$ and $\bSigma_{k,l}$ are unitary and diagonal matrices, respectively. Denote the diagonal elements of $\bSigma_{k,l}=\diag{\vartheta_{k,l,1},...,\vartheta_{k,l,N_k}}$, where $\vartheta_{k,l,1}\geq...\geq\vartheta_{k,l,N_k}\geq 0$. Then, the optimal tramsmit covariance matrix in user $k$ at layer $l$, i.e., $\bQ^{(\ell),[t+1]}_{k,l}$, can be expressed as
	\begin{align}\label{equ:wfsol}
	\bQ^{(\ell),[t+1]}_{k,l}=(\bS_{k,l}^{\star})^{-\frac{1}{2}}\widetilde{\bU}_{k,l}\bLambda_{k,l}\widetilde{\bU}_{k,l}^H(\bS_{k,l}^{\star})^{-\frac{1}{2}},
	\end{align}
	where
	\begin{align}\label{equ:Lambda}
	\bLambda_{k,l}=\diag{\left(1-\frac{1}{\vartheta_{k,l,n}}\right)^{+}}_{n=1}^{N_k}.
	\end{align}
\end{prop}
\IEEEproof Refer to \appref{app:B}.

Note that the DE parameters $\{\bGamma_{k,l}\}_{\forall k,l}$ in \eqref{equ:SGammaDecom} are also dereived iteratively with given optimal solutions $\{\bQ_{k,l}^{\star}\}_{\forall k,l}$ of problem \eqref{eq:problem5}. Therefore, we resort to the alternating optimization (AO) approach to optimize between $\{\bGamma_{k,l}\}_{\forall k,l}$ and $\{\bQ_{k,l}\}_{\forall k,l}$, where we cyclically calculating $\{\bQ_{k,l}\}_{\forall k,l}$ by \emph{\propref{Prop:Water-Filling}} with given $\{\bGamma_{k,l}\}_{\forall k,l}$, and then update $\{\bGamma_{k,l}\}_{\forall k,l}$ through \textbf{\alref{alg:DEs}} with the updated $\{\bQ_{k,l}\}_{\forall k,l}$. Finally, a modified EM exposure-aware water-filling scheme is proposed in \textbf{\alref{alg:waterfilling}}. 
\begin{algorithm}[h]
	\caption{EM Exposure-Aware Water-Filling Algorithm}
	\label{alg:waterfilling}
	\begin{algorithmic}[1]
		\Require Initial dual variables $\{\mu_k^{(0)}\}_{\forall k}$ and $\{\lambda_{k,a}^{(0)}\}_{\forall k,a}$, feasible $\left\{\bQ_{k,l}^{(0)}\right\}_{\forall k,l}$ and $\bpi$, channel statistics $\left\{\bOmega_{k}\right\}_{\forall k}$, necessary constants in problem \eqref{eq:problem5}, iteration thresholds $\epsilon_4$.
		\Ensure The optimal solution $\left\{\bQ_{k,l}^{\star}\right\}_{\forall k,l}$ of $\cP_5^{(\ell),[t]}$. 
		\State Initialize the iteration indices $v_1=0$, $v_2=0$.
		\Repeat
		\State Get $\bS_{k,l}^{(v_1)},\forall k,l$ by \eqref{equ:Skl} with $\mu_k^{(v_1)}$ and $\lambda_{k,a}^{(v_1)},\forall k,a$.
		\State Set $\left\{\bQ_{k,l,(0)}^{(v_1)}\right\}_{\forall k,l}=\left\{\bQ_{k,l}^{(v_1)}\right\}_{\forall k,l}$ and $v_2=0$.
		\Repeat
		\State Obtain DE parameters $\bGamma_{k,l,(v_2)},\forall k,l$ by \textbf{\alref{alg:DEs}} with given $\bQ_{k,l,(v_2)}^{(v_1)},\forall k,l$.
		\State  Calculate $\widetilde{\bU}_{k,l,(v_2)}$ and $\bSigma_{k,l,(v_2)},\forall k,l$ by \eqref{equ:SGammaDecom}.
		\State Obtain $\bQ_{k,l,(v_2+1)}^{(v_1)},\forall k,l$ by \eqref{equ:wfsol} and \eqref{equ:Lambda}.
		\State Set $v_2=v_2+1$.
		\Until{$\left|\etaEEb(\bQ^{(v_1)}_{(v_2)})-\etaEEb(\bQ^{(v_1)}_{(v_2-1)})\right|<\epsilon_4$.}
		\State Return $\left\{\bQ_{k,l}^{(v_1+1)}\right\}_{\forall k,l}=\left\{\bQ_{k,l,(v_2)}^{(v_1)}\right\}_{\forall k,l}$.
		\State Set $v_1=v_1+1$.
		\State Update the dual variables $\mu_k^{(v_1)}$ and $\lambda_{k,a}^{(v_1)},\forall k,a$ by minimizing $\mathcal{LA}(\bQ^{(v_1)},\{\mu_k\},\{\lambda_{k,a}\})$ in \eqref{equ:dual_problem}.
		\Until{The dual variables $\{\mu_k\}_{\forall k}$ and $\{\lambda_{k,a}\}_{\forall k,a}$ converge.}
		\State Return $\left\{\bQ_{k,l}^{\star}\right\}_{\forall k,l}=\left\{\bQ_{k,l}^{v_1}\right\}_{\forall k,l}$.
	\end{algorithmic}
\end{algorithm}
Then, substituting this modified EM exposure-aware water-filling algorithm for step $7$ of \textbf{\alref{alg:MMDinkel}}, the approximate solutions of $\cP_2^{\mathrm{in}}$ can be obtained at low complexity by performing all the steps of \textbf{\alref{alg:MMDinkel}} with a given permutation of decoding order $\bpi$.

\subsection{Decoding Order Optimzation}\label{subsec:Decoding}
In the previous subsections, we have investigated the EM exposure EE maximization design for transmit covariance matrices $\{\bQ_{k,l}\}_{\forall k,l}$ with fixed permutation $\bpi$, and obtained the maximal $\etaEE(\bQ)$, which is also written as $f(\bpi)$ in the problem \eqref{eq:problemlow}. Now, we also need to optimize $\bpi$ in the outer level, where the problem is formulated as \eqref{eq:problemup}. In fact, the permutation $\bpi$ determines $f(\bpi)$ by determining $\bK_{k,l}$ in \eqref{equ:Kkl}. The optimization of decoding order can effectively alleviate the interference from other users or layers of the decoded signal in the SIC procedure so that the overall rate throughput and the system EE can be improved significantly. 

Note that handling $\cP_2^{\mathrm{out}}$ is difficult because of the non-convex optimization function and the discrete variable $\bpi$. For problem \eqref{eq:problemup}, the optimal permutation $\bpi$ can be obtained by exhaustive search, which ensures the optimal solution by selecting the decoding order that maximizes the objective function from the possible discrete value space $\bPi$. However, the complexity of this method will grow more than exponentially when the users and layers increase constantly, which is prohibitive for the overall algorithm. Specifically, the optimization in our considered system with $K$ users, where the transmit signal of each user is split into $L$ layers, requires a search over $(KL)!$ permutations of user and layer indices. To deal with this challenge, we propose a low complexity scheme considering only one permutation of all users and layers.

Similar to the optimization of decoding order in NOMA systems, we arrange the permutation $\bpi$ according to the rank of user $k$ in terms of its maximum sum-rate in the single user system \cite{KS21}. Specifically, denote that $R_k^{\mathrm{SU}},\forall k$ as the maximum sum rate of users $k\in\mathcal{K}$ in single user systems, respectively, which can be formulated by
\begin{subequations}\label{eq:tanxin}
	\begin{align}
	R_k^{\mathrm{SU}}=\underset{\bQ_{k}\succeq\mathbf{0}} \max &\ \mathbb{E}\left\{\log\det\left(\bI_{M}+\frac{1}{\sigmatwo}\bH_k\bQ_{k}\bH_k^H\right)\right\}, \label{pta}\\
	{\mathrm{s.t.}} &\ \tr{\bQ_{k}}\leq\Pmaxk, \label{ptb}\\
	&\ \tr{\bR_{k,a}\bQ_{k}}\leq D_{k,a}, \quad\forall k,a, \label{ptc}
	\end{align}
\end{subequations}
where $\bQ_k\triangleq\mathbb{E}\{\bx_k\bx_k^H\}=\sum_{l=1}^{L}\bQ_{k,l},\forall k$ is the transmit covariance matrix of user $k$. Then, our optimization of the user index order $\left(k_1,k_2,...,k_K\right)$ in problem \eqref{eq:problemup} satisfies $R_{k_1}^{\mathrm{SU}}>R_{k_2}^{\mathrm{SU}}>...>R_{k_K}^{\mathrm{SU}}$, where $k_1,k_2,...,k_K\in\mathcal{K}$ denote the indices of users in our considered system. Under this principle, the layers that belong to the same user share the same optimization expression, which are exchangeable in the decoding order. Without losing generality, the approximate solution of decoding order of SIC can be written as
\begin{align}\label{equ:decodingorder}
\bpi=\left(\pi_{k_1,1},...,\pi_{k_1,L},\pi_{k_2,1},...\pi_{k_2,L},...,\pi_{k_K,1},...,\pi_{k_K,L}\right).
\end{align} 

Note that \eqref{equ:decodingorder} utilizes the idea of the greedy algorithm so that it can achieve the local optimum of $\cP_6$. More importantly, the complexity of handling problem \eqref{eq:problemup} is greatly reduced at a very slight cost, especially for the RSMA systems with a lot of users and layers. With this in mind, the overall EM exposure-aware EE maximization algorithm for multiuser massive MIMO uplink RSMA is proposed in \textbf{\alref{alg:Overall}}.

\begin{algorithm}[h]
	\caption{Overall EE Maximization Algorithm for Uplink RSMA with EM Exposure Constraints}
	\label{alg:Overall}
	\begin{algorithmic}[1]
		\Require Channel statistics $\left\{\bOmega_{k}\right\}_{\forall k}$, determined set $\bPi$.
		\Ensure Approximate solutions $\left\{\bQ_{k,l}^{\mathrm{opt}}\right\}_{\forall k,l}$ and $\bpi^{\mathrm{opt}}$ of $\cP_1$.
		\State Obtain $R_k^{\mathrm{SU}}, \forall k$ by solving the problem \eqref{eq:tanxin}.
		\State Arrange $R_k^{\mathrm{SU}}, \forall k$ by descending order and obtain the permutation $\bpi$ by \eqref{equ:decodingorder}.
		\State Deterimine $\mathcal{\bm{Q}}_{k,l}^{\bpi}$ and $\bK_{k,l}, \forall k,l$ with given $\bpi$.
		\State Obtain the approximate solution $\left\{\bQ_{k,l}^{\bpi}\right\}_{\forall k,l}$ of $\cP_2^{\mathrm{in}}$ by \textbf{\alref{alg:MMDinkel}} and \textbf{\alref{alg:waterfilling}} with given $\mathcal{\bm{Q}}_{k,l}^{\bpi}$ and $\bK_{k,l}, \forall k,l$.
	\end{algorithmic}
\end{algorithm}

\subsection{Convergence and Complexity Analysis}
In this sub-section, we intend to analyze the convergence and complexity of the proposed EM exposure EE maximization algorithm for uplink RSMA. Due to the fact that the result of \emph{\propref{Prop:Water-Filling}} is essentially the solution of \eqref{equ:dual_problem}, which is the strong dual problem of the convex problem $\cP_5^{(\ell),[t]}$, the water-filling scheme iterating the dual variables $\{\mu_k\}_{\forall k}$ and $\{\lambda_{k,a}\}_{\forall k,a}$ in \textbf{\alref{alg:waterfilling}} will certainly converges to the global optimum of the problem \eqref{eq:problem5} \cite{BV04}. In addition, \textbf{\alref{alg:MMDinkel}} tackles the problem \eqref{eq:problem3} by utilizing the MM method and Dinkelbach's transform, where its convergence is guaranteed according to \cite{SBP17,ZJ15}. Generally, \textbf{\alref{alg:Overall}} optimizes the decoding order in advance and then maximizes EE with given permutation $\bpi$ by \textbf{\alref{alg:MMDinkel}} and \textbf{\alref{alg:waterfilling}}. Therefore, after determining the permutation by the greedy approach, $\cP_1$ can be finally handled by \textbf{\alref{alg:Overall}}, which will ensure the approximate solutions of $\cP_1$.

Then, we analyze the computational complexity of the proposed overall algorithm in detail. Note that the complexity of the iterative algorithm depends on the number of iterations, which can be dynamically adjusted by inputing different thresholds.  
In \textbf{\alref{alg:waterfilling}}, the internal iteration including steps 5--10 corresponds to the computation of \emph{\propref{Prop:Water-Filling}} with given dual variables, where the main complexity locates in \eqref{equ:SGammaDecom} and can be expressed as $\mathcal{O}(\sum_{k}LN_k^3)$. 
Consider that $\wbTheta(\bX)$ and $\bTheta(\bX)$ are diagonal matrices, the calculation of DE parameters in \eqref{equ:DEpara1}--\eqref{equ:DEpara4} only requires a few linear operations, where the complexity is estimated as $\mathcal{O}(\sum_{k}LN_k^2)$. Because of the fast convergence of DE parameters as indicated in [31], the complexity of \textbf{\alref{alg:DEs}} can be neglected compared with that of \eqref{equ:SGammaDecom}.
In addition, the external iteration of \textbf{\alref{alg:waterfilling}} aims to find the optimal dual variables by minimizing the Lagrange function at step 13. Due to the convexity of the dual problem, the complexity of step 13 is expressed by $\mathcal{O}\left((K+\sum_{k=1}^KA_k)^x\right)$, where $1\leq x\leq 4$ is determined by the convex program \cite{NCN21}. Assume that the numbers of internal and external iterations of \textbf{\alref{alg:waterfilling}} are $I_1$ and $I_2$, respectively. Then, the complexity of \textbf{\alref{alg:waterfilling}} can be
estimated as $\mathcal{O}\left(I_2\left(I_1\sum_{k}LN_k^3+(K+\sum_{k=1}^KA_k)^x\right)\right)$. Moreover, \textbf{\alref{alg:MMDinkel}} consists of the outer MM and inner Dinkelbach's approaches, which are assumed to perform $I_3$ and $I_4$ iterations, respectively. The complexity of \textbf{\alref{alg:MMDinkel}} depends mainly on tackling the inner layer problem, i.e., step 7, which is described by \textbf{\alref{alg:waterfilling}}. 
Furthermore, the complexity of tackling \eqref{eq:tanxin} would be similar to the complexity of tackling \eqref{eq:problem5}. Then, the complexity of step 1 of \textbf{\textbf{\alref{alg:Overall}}} is comparable to \textbf{\alref{alg:MMDinkel}} without the outermost two iterations. Therefore, the overall computational complexity of \textbf{\textbf{\alref{alg:Overall}}} is approximately $\mathcal{O}\left((I_4I_3+1)I_2\left(I_1\sum_{k}LN_k^3+(K+\sum_{k=1}^KA_k)^x\right)\right)$.

\section{Numerical Results}\label{sec:numerical_results}
In this section, the performance of the proposed EM exposure-aware EE maximization design for multiuser uplink RSMA transmission is appraised through extensive computer simulations. 
Consider the suburban scenario with non-line-of-sight (NLOS) propagation, and the BS is equipped with half-wavelength antenna spacing ULA. In this case, DFT matrices can well approximate the deterministic unitary matrices in Weichselberger’s channel decomposition model \cite{YGX15,YGS16}. Therefore, the values of $\bU$ and $\bV_k, \forall k$ are set to DFT matrices in our simulation \cite{S05}. 
Suppose that the system bandwidth is $W=10$ MHz, the number of SAR constraints at each user is $A_k=1$, the number of antennas at each user is $N_k=4$ and the number of antennas at the BS is $M=64$, the noise covariance is $\sigmatwo=-96$ dBm, the path loss is 120 dB, the amplifier inefficiency is $1/\xi_k = 0.2, \forall k$, the power dissipations at the users and BS are $P_{c,k}=30$ dBm, $\forall k$ and $P_{\mathrm{BS}}=40$ dBm, respectively.
In our simulations, the number of users is $K=4$, and the signal from each user is split into two layers, which means $L=2$, unless specified otherwise. Then, the permutation $\bpi$ is composed of $K\cdot L=8$ elements.

For the clarity of the simulation results, we assume that the power constraints and the SAR constraints are the same for all users, as adopted in \cite{YLH15,YLH17}. Then, the power and SAR budgets in our simulations can be written as $\Pmaxk=\Pmax, \forall k$ and $D_{k,a}=D=0.8\ \mathrm{W/kg}, \forall k,a$. Typically, the SAR matrix can be obtained numerically by fitting measurements from a phantom for human head with the given gesture \cite{HLY13}. In the following, we take the SAR matrix adopted in \cite{YLH17} as an example to show the performance of the proposed algorithm, which is given by
\begin{align}
\Rka=\bR=
\left[
\begin{array}{cccc}
8 &-6\jmath &-2.1 &0 \\
6\jmath &8 &-6\jmath &-2.1 \\
-2.1 &6\jmath &8 &-6\jmath \\
0 &-2.1 &6\jmath &8
\end{array}
\right], \forall k,a.
\end{align}

%\begin{figure}[t]
%	\centering
%	\subfloat[]{\centering\includegraphics[width=0.35\textwidth]{EEiterIone_plus.eps}\label{fig:EEiterIone}}
%	\hfill
%	\subfloat[]{\centering\includegraphics[width=0.35\textwidth]{EEiterItwo_plus.eps}\label{fig:EEiterItwo}}
%	\hfill
%	\subfloat[]{\centering\includegraphics[width=0.35\textwidth]{EEiterIthree_plus.eps}\label{fig:EEiterIthree}}
%	\hfill
%	\subfloat[]{\centering\includegraphics[width=0.35\textwidth]{EEiterIfour_plus.eps}\label{fig:EEiterIfour}}
%	\caption{The convergence performance of the proposed EM exposure-aware EE maximization algorithm under different power budgets $\Pmax$. (a) $I_1$;  (b) $I_2$; (c) $I_3$; (d) $I_4$.}
%	\label{fig:EEiter}
%\end{figure}

\begin{figure}
	\begin{minipage}[t]{0.495\linewidth}
		\centering
		\subfloat[]{\centering\includegraphics[width=1.0\textwidth]{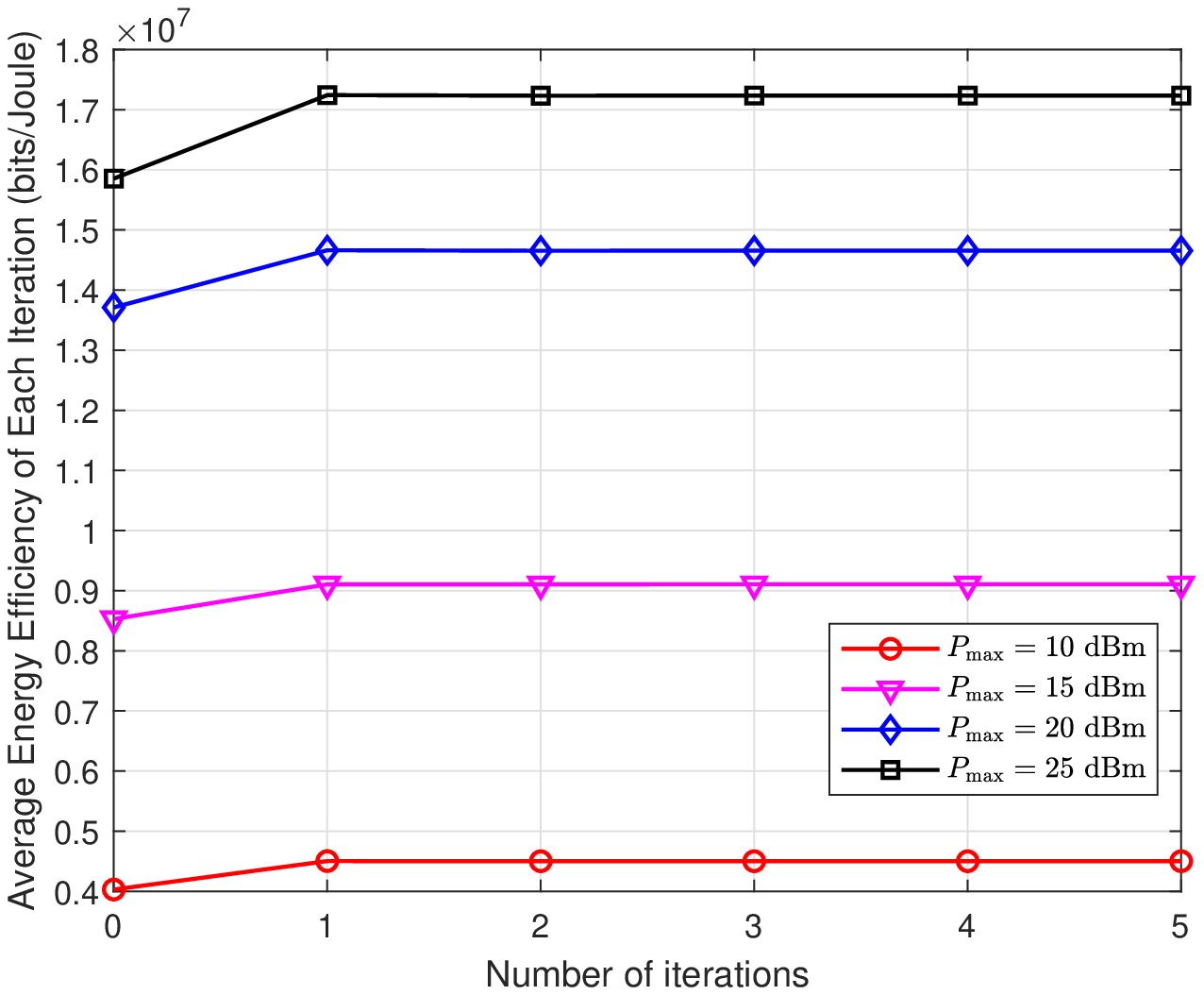}\label{fig:EEiterIone}}
	\end{minipage}%
	\hspace{0.0cm}
	\begin{minipage}[t]{0.495\linewidth}
		\centering
		\subfloat[]{\centering\includegraphics[width=1.0\textwidth]{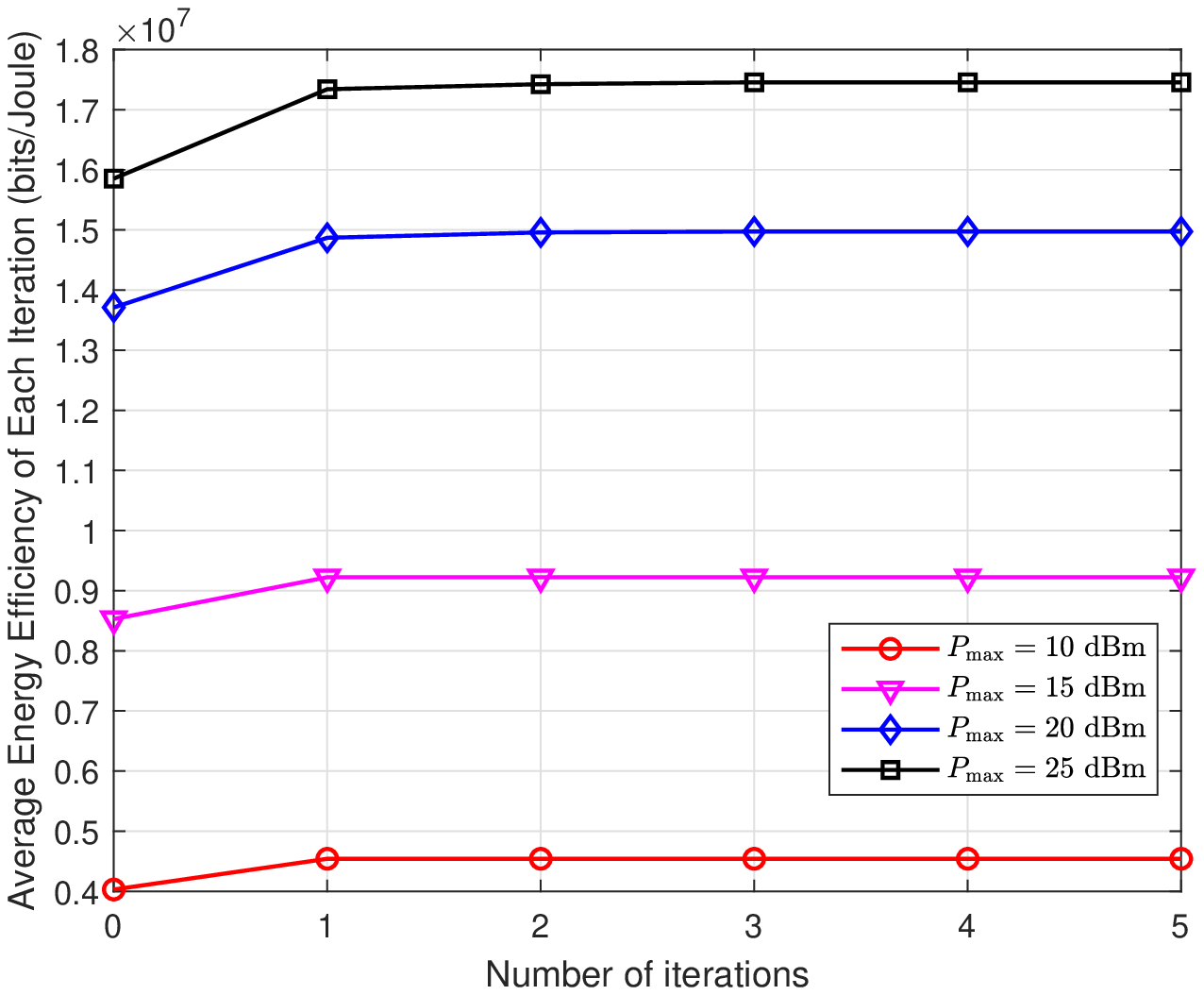}\label{fig:EEiterItwo}}
	\end{minipage}
%		\hfill
	\begin{minipage}[t]{0.495\linewidth}
		\centering
		\subfloat[]{\centering\includegraphics[width=1.0\textwidth]{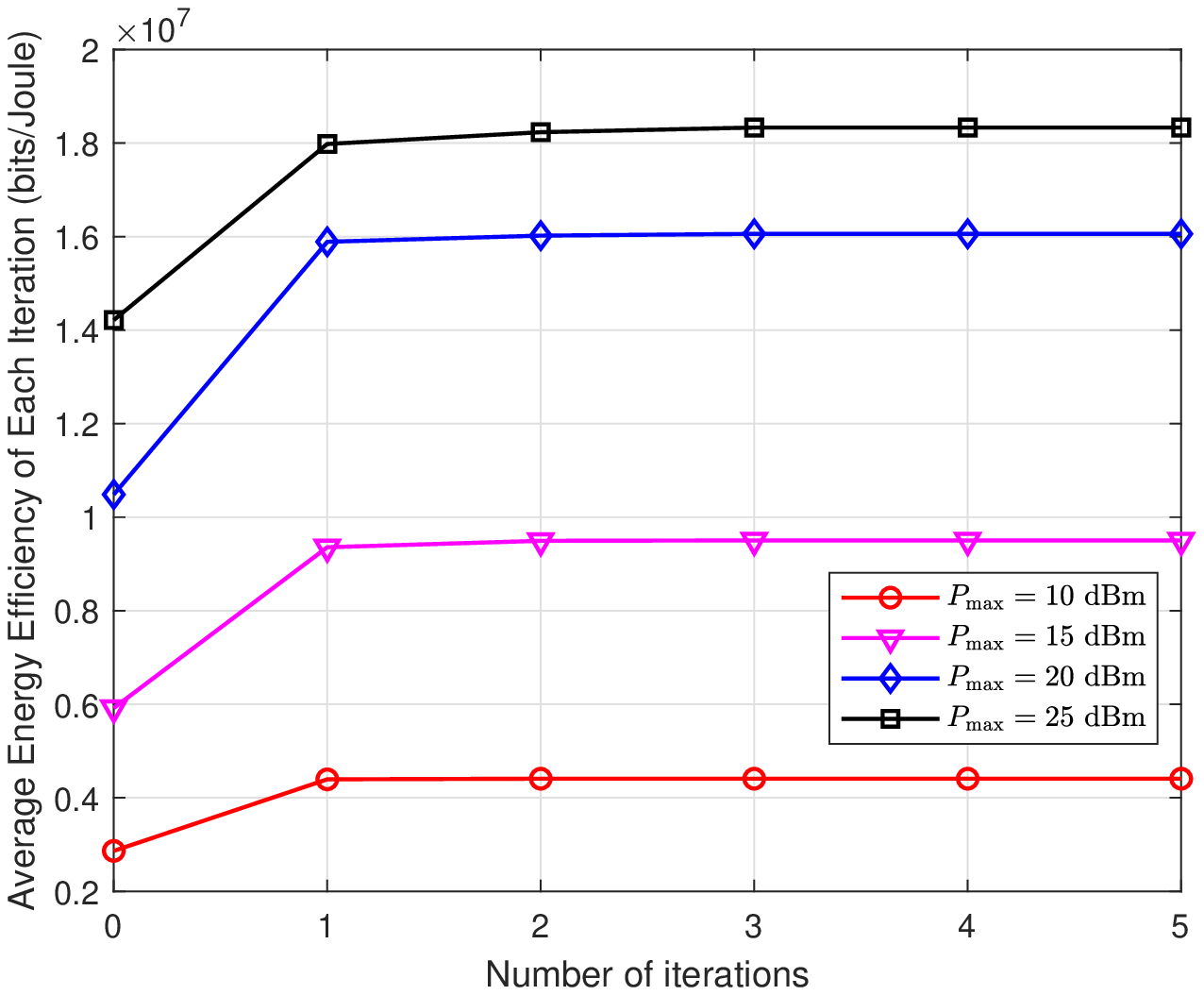}\label{fig:EEiterIthree}}
	\end{minipage}%
	\hspace{0.0cm}
	\begin{minipage}[t]{0.495\linewidth}
		\centering
		\subfloat[]{\centering\includegraphics[width=1.0\textwidth]{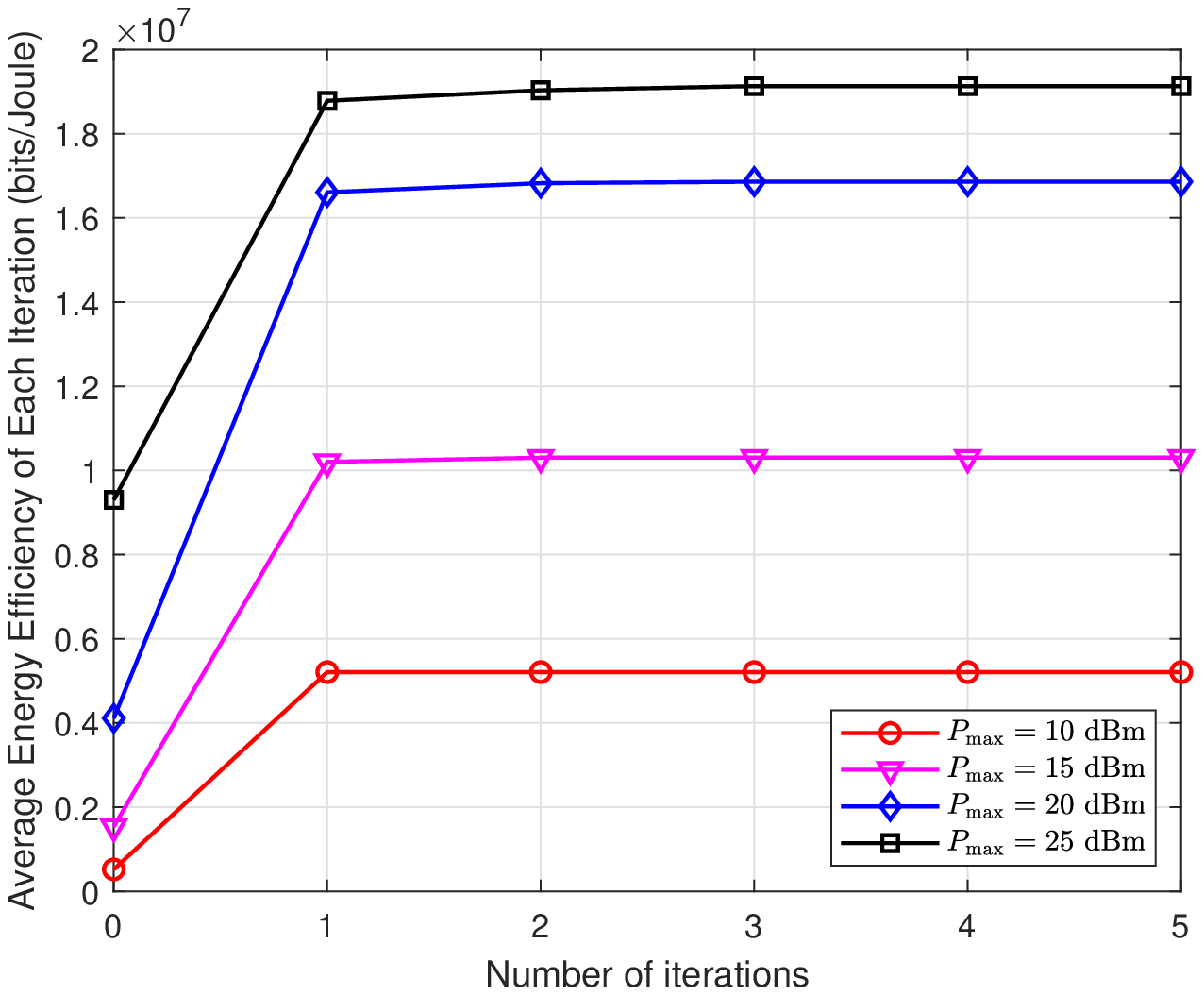}\label{fig:EEiterIfour}}
	\end{minipage}
	\caption{The convergence performance of the proposed EM exposure-aware EE maximization algorithm under different power budgets $\Pmax$. (a) $I_1$;  (b) $I_2$; (c) $I_3$; (d) $I_4$.}
	\label{fig:EEiter}
\end{figure}

\figref{fig:EEiter} describes the convergence performance versus iterations $I_1$--$I_4$ of the proposed algorithm under different power budgets, where we record the average EE values obtained after each iteration of the overall algorithm. The results indicate that all the iterations of the proposed algorithm exhibit fast convergence rates. In particular, in the region of low $\Pmax$, the overall algorithm can converge with only one iteration.

\begin{figure}[t]
		\centering
		\includegraphics[width=0.45\textwidth]{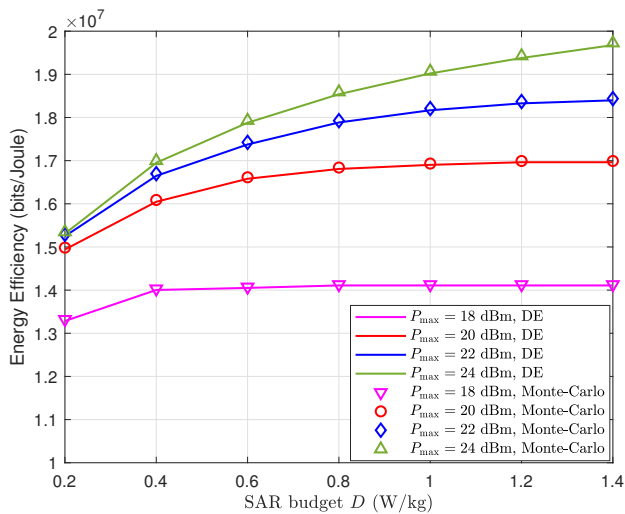}
		\caption{The EE performance versus SAR budgets under different power constraints: DE and Monte-Carlo methods.}
		\label{fig:eeDdifp3}
\end{figure}

\begin{figure}[t]
		\centering
		\includegraphics[width=0.45\textwidth]{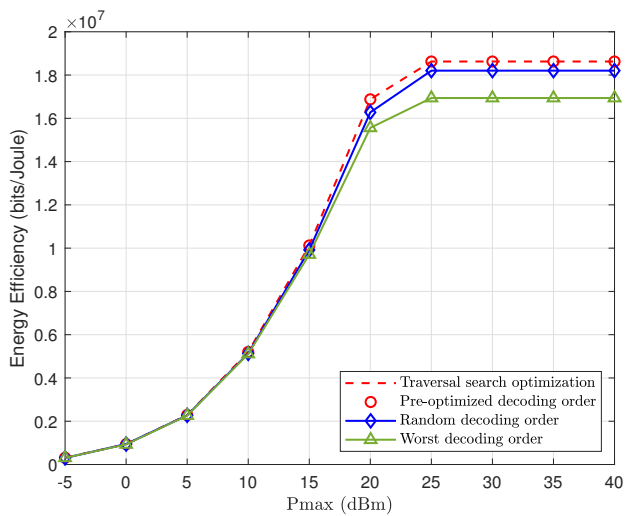}
		\caption{The EE performance of the proposed optimization approaches for different decoding orders ($D=0.8\ \mathrm{W/kg}$).}
		\label{fig:eepOptrandom2}
\end{figure}

Nextly, the EE performance for different EM exposure constraints is presented in \figref{fig:eeDdifp3}. It can be observed that the increase in the SAR budgets will improve the EE performance of the system in the region of high $\Pmax$. When $\Pmax$ is relatively low, the power constraints provide the main limiting condition for system EE compared with the SAR constraints. On this condition, the relaxation of SAR constraints has no effect on system EE. However, when $\Pmax$ is large enough, SAR has gradually become the main constraint of the optimization problem, which has a decisive impact on the maximum EE. Therefore, the EE curves of \figref{fig:eeDdifp3} with large power budgets can still monotonically increase under the high SAR budget regime. In addition, Fig. 3 also shows the EE performance gap between the DE and Monte-Carlo methods. In the simulations, we calculate the average EE by the Monte-Carlo method with 1000 channel matrices samples. From the results, it can be noted that the DE method provides a close approximation to the ergodic EE of uplink RSMA systems.

\figref{fig:eepOptrandom2} shows the EE performance gap caused by different optimization methods of decoding order $\bpi$. The first method refers to traversing all the feasible sets of decoding order $\bPi$ to search for the optimal permutation that can maximize the system EE. The second method is to determine the decoding order in advance based on the channel characteristics of different users, which is indicated in \eqref{eq:tanxin} and \eqref{equ:decodingorder}. In addition, we present the maximum EE performance under random decoding order and worst decoding order, which refers to the decoding order opposite to the optimal permutation, as shown by the blue and green curves of Fig. 4, respectively. The result shows that the optimization of decoding order permutation improves the system EE. 
Besides, the performance loss incurred by the greedy approach is shown to be negligible compared with the exhaustive approach.
Since the multiple access interference is only related to the receive signals of the layers that are decoded later, the greedy approach, where the users and layers with large channel gain are decoded previously, may minimize the interference during the decoding. Therefore, the performance gap between the proposed and second method can be small.

\begin{figure}[t]
		\centering
		\includegraphics[width=0.45\textwidth]{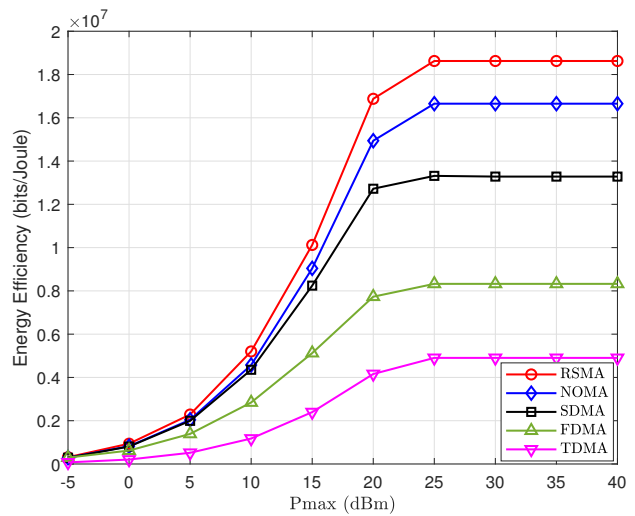}
		\caption{Comparision of EE performance among RSMA, NOMA, SDMA, FDMA and TDMA transmission schemes.}
		\label{fig:eepRSMA2}
\end{figure}%

\begin{figure}[t]
		\centering
		\includegraphics[width=0.45\textwidth]{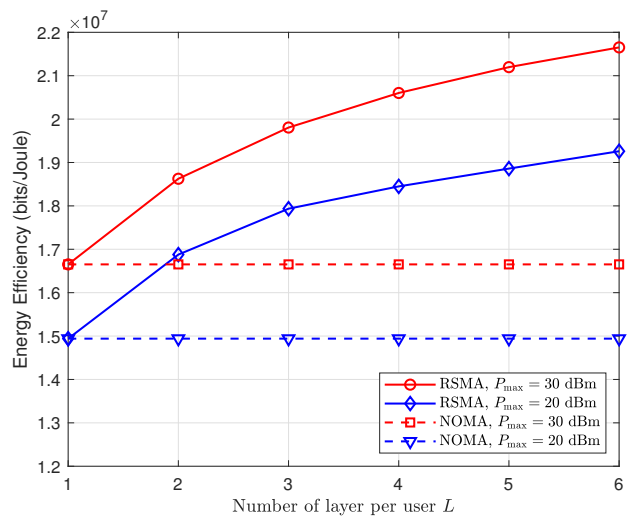}
		\caption{Impact of the number of layers, $L$, on the maximum achievable EE performance.}
		\label{fig:eeL}
\end{figure}

Moreover, for the purpose of appraising the performance gain of uplink RSMA transmission in the criterion of EE maximization, we intend to investigate the EM exposure-aware EE maximization problems for uplink NOMA, space division multiple access (SDMA), frequency division multiple access (FDMA), and time division multiple access (TDMA) systems, respectively. For comparison purposes, we provide the problem formulations under the above baseline schemes with both power and SAR constraints, and then propose the corresponding algorithms to deal with these problems. Note that due to the introduction of SAR constraints, the EM exposure-aware optimizations for these uplink baseline transmissions are projected as a point that has not
been studied in the literature.  

\begin{itemize}
	\item \emph{$\mathbf{NOMA}$}: In the uplink NOMA transmissions, the BS first decodes the signals of stronger users, and then decodes the remaining signals by subtracting interference from the decoded stronger user \cite{IAD17}. 
	Then, the ergodic rate of user $k$ under the NOMA scheme is written as
	\begin{align}\label{equ:rateNOMA}
	&R_{k}^{\mathrm{NOMA}}=\mathbb{E}\Bigg\{\log\det\Bigg(\bI_{M}+\ntb
	&(\sigmatwo\bI_{M}+\sum_{z=k+1}^{K}\bH_z\bQ_{z}\bH_z^H)^{-1}\bH_k\bQ_{k}\bH_k^H\Bigg)\Bigg\}.
	\end{align}
	
	\item \emph{$\mathbf{SDMA}$}: As indicated by \cite{MCL18}, in the SDMA scheme, the BS decodes the desired message by treating other interference as noise. Then, the ergodic rate of user $k$ can be expressed as 
	\begin{align}\label{equ:rateSDMA}
	&R_{k}^{\mathrm{SDMA}}=\mathbb{E}\Bigg\{\log\det\Bigg(\bI_{M}+\ntb
	&(\sigmatwo\bI_{M}+\sum_{k^{\prime}\neq k}\bH_{k^{\prime}}\bQ_{k^{\prime}}\bH_{k^{\prime}}^H)^{-1}\bH_k\bQ_{k}\bH_k^H\Bigg)\Bigg\}.
	\end{align}
	
	\item \emph{$\mathbf{FDMA}$}: In the FDMA scheme, the transmit signal of each user occupies a fraction of the total bandwidth $W$. Denote $\alpha_k$ as the bandwidth fraction allocated to user $k$, where $0\leq\alpha_k\leq 1$ and $\sum_{k=1}^{K}\alpha_k=1$. Assume that users are allocated the same bandwidth fraction, i.e., $\alpha_k=1/K,\forall k$. Then, the covariance of additive Gaussian noise of user $k$ can be expressed as $\sigmatwo/K$. At the BS, the ergodic rate of user $k$ can be written as
	\begin{align}\label{equ:rateFDMA}
	R_{k}^{\mathrm{FDMA}}=\mathbb{E}\Bigg\{\frac{1}{K}\log\det\Bigg(\bI_{M}+\frac{K}{\sigmatwo}\bH_k\bQ_{k}\bH_k^H\Bigg)\Bigg\}.
	\end{align}
	
	\item \emph{$\mathbf{TDMA}$}: In the TDMA scheme, each user is assigned a fraction of time to use the whole bandwidth. Denote $\beta_k$ as the time fraction allocated to user $k$, where $0\leq\beta_k\leq 1$ and $\sum_{k=1}^{K}\beta_k=1$. Assume that users occupy the same time fraction, i.e., $\beta_k=1/K,\forall k$. Then, the ergodic rate of user $k$ can be written as
	\begin{align}\label{equ:rateTDMA}
	R_{k}^{\mathrm{TDMA}}=\mathbb{E}\Bigg\{\frac{1}{K}\log\det\Bigg(\bI_{M}+\frac{1}{\sigmatwo}\bH_k\bQ_{k}\bH_k^H\Bigg)\Bigg\}.
	\end{align}
	
\end{itemize}
Utilizing the definition of EE in \eqref{equ:EE}, the energy consumption can be rewritten as $P(\bQ)=\sum_{k=1}^{K}\left(\xi_{k}\tr{\bQ_k}+P_{c,k}\right)+P_{\mathrm{BS}}$. Then, the EM exposure-aware EE maximization problem for uplink NOMA, FDMA and TDMA can be expressed as
\begin{subequations}\label{eq:problemNOMA}
	\begin{align}
	\cP^{\mathrm{UL}}:\ \underset{\bQ_k\succeq \mathbf{0},\forall k} \max &\  \mathrm{EE}^{\mathrm{UL}}=\sum\limits_{k=1}^{K}R_{k}/P(\bQ), \label{pNOMAa}\\
	{\mathrm{s.t.}} &\  \tr{\bQ_{k}}\leq\Pmaxk, \label{pNOMAb}\\
	&\ \tr{\bR_{k,a}\bQ_{k}}\leq D_{k,a}, \quad\forall k,a, \label{pNOMAc}
	\end{align}
\end{subequations}
where $R_k$ is taken as $R_{k}^{\mathrm{NOMA}}$, $R_{k}^{\mathrm{FDMA}}$ and $R_{k}^{\mathrm{TDMA}}$ in \eqref{equ:rateNOMA}, \eqref{equ:rateFDMA} and \eqref{equ:rateTDMA}, respectively.

\figref{fig:eepRSMA2} compares the maximum EE performance among RSMA and three baseline transmission schemes. Actually, the main difference in EE of the four multiple access modes mainly comes from their transmission SE performances. Compared with TDMA, the FDMA uses a narrower frequency band to transmit the data of each user and layer, thus resulting in the reduction of noise energy and improving the system SE. 
Furthermore, within the same bandwidth, the SDMA scheme can realize the parallel data transmission of multiple users, while FDMA can only transmit the data stream of one user. Therefore, the system SE of SDMA can be greatly improved compared with that of FDMA. Since the NOMA scheme has lower multiple access interference than SDMA scheme, it can achieve better EE performance at the cost of enhancing the complexity of the receiver. 
Compared with NOMA, RSMA continues to split each non-orthogonal user data stream in the same time-frequency resource into multiple layers and then independently transmit them in parallel, which further improves the system SE. Therefore, when the number of layers per user exceeds one, RSMA can achieve higher EE than the NOMA transmission scheme.

Actually, the improvement of EE performance of RSMA compared with NOMA is also related to the number of layers. \figref{fig:eeL} presents the impact of the number of layers, $L$, on the maximum achievable EE in the RSMA transmission scheme. As expected, the EE performance increases as the number of layers grows. 
However, due to the existence of multiple access interference, the improvement of system EE by the number of layers is not linear. When there are too many layers divided by each user, the interference of each layer decoded by the receiver will increase correspondingly, which limits the improvement of system EE. Note that $L$ cannot grow boundless in practical scenarios. The number of RSMA layers is limited by the data flow and the decoding complexity at the receiver. Meanwhile, the increase of layers demands higher complexity of decoding at the receiver, thus reducing the efficiency of the algorithm. Generally, it would be best to select the proper number of split layers in actual RSMA systems.

\begin{figure}[t]
		\centering
		\includegraphics[width=0.45\textwidth]{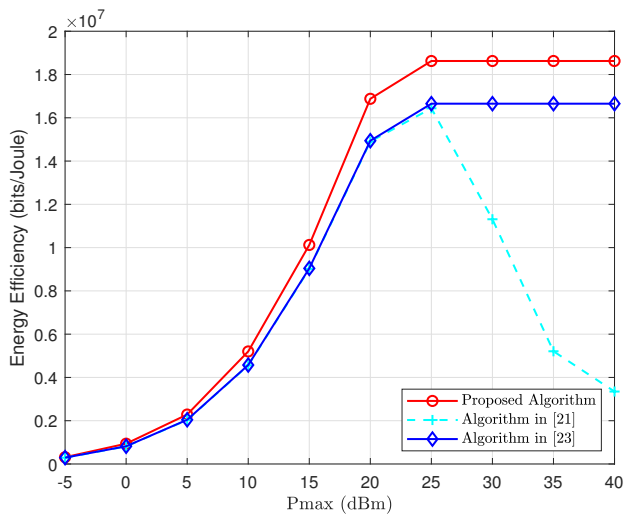}
		\caption{Comparison of EE performance between the proposed algorithm and the algorithms in \cite{YLH17} and \cite{XYN21}.}
		\label{fig:comparision}
\end{figure}%
\begin{figure}[t]
		\centering
		\includegraphics[width=0.45\textwidth]{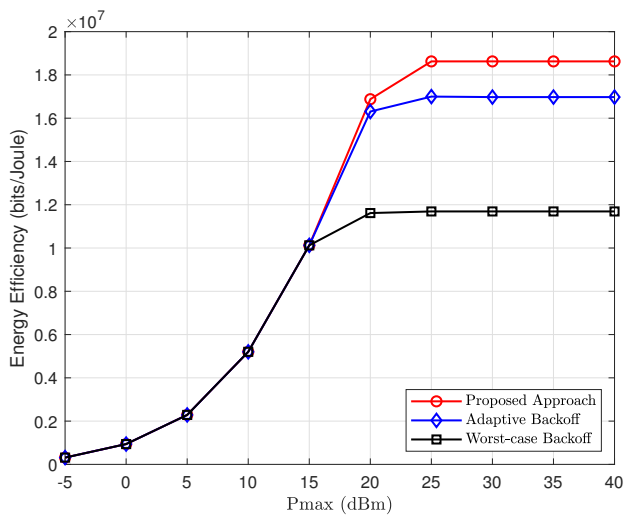}
		\caption{Comparison of EE performance between the proposed and baseline approaches.}
		\label{fig:eepbackoff}
\end{figure}

\figref{fig:comparision} compares the EE performance of the proposed algorithm in this paper with that in \cite{YLH17} and \cite{XYN21}. Note that work in \cite{YLH17} and \cite{XYN21} investigated the sum-rate and EE maximization for uplink multiuser MIMO systems with SAR constraints, respectively. It can be observed that the algorithm proposed in this paper has superior EE performance. It is worth noting that if we optimize the transmit covariance matrices in the criterion of sum-rate maximization, the system EE will decrease under high power budget. This is because when $\Pmax$ is high, the power budget is fully used to achieve high sum-rate, which also increases the total power consumption of the considered system and leads to the reduction of EE. Basically, the algorithm proposed in this paper is more general compared with that in \cite{YLH17} and \cite{XYN21}. In fact, the EE maximization problem studied in \cite{XYN21} can be regarded as a special case of the problem in this paper by setting the number of layers $L$ to one. For the algorithm in \cite{YLH17}, we can further set $\xi_k=0, \forall k$ to achieve the same function.

To demonstrate the effectiveness of the proposed EM exposure-aware active design, \figref{fig:eepbackoff} compares the EE performance of the proposed algorithm with baseline approaches, i.e., the adaptive and worst-case power backoff approaches, which only consider the power constraints in the criterion of EE maximization and then attenuate the transmit power to satisfy the SAR constraints. Specifically, we assume that \eqref{plowc} is not taken into account in the optimization problem $\cP_2^{\mathrm{in}}$. Then the two baseline approaches can make the EM exposure non-aware results of $\cP_2^{\mathrm{in}}$ satisfy the condition $\eqref{plowc}$ by introducing the backoff factor $\alpha$.
% where the detailed process is expressed as follows:
\begin{itemize}
	\item \emph{Adaptive Backoff} \cite{YLH15}: Consider the problem \eqref{eq:problemlow} that omits the SAR constraints, i.e.,
	\begin{align}\label{equ:Adpbackoff1}
	\bQ^{\mathrm{0}}=\argmax{\sum_{l=1}^{L}\tr{\bQ_{k,l}}\leq\Pmaxk}\ \mathrm{EE}(\bQ,\bpi), \quad \forall k.
	\end{align}
	Then, the adaptive backoff method makes the final result satisfy the SAR constraints by altering the transmit covariances as
	\begin{align}\label{equ:Adpbackoff2}
	\bQ^{\mathrm{adp}}_{k,l,\mathrm{opt}}=\alpha_{k}^{\mathrm{adp}}\bQ_{k,l}^{\mathrm{0}},\quad \forall k,l,
	\end{align}
	where
	\begin{align}\label{equ:Adpbackoff3}
	\alpha_{k}^{\mathrm{adp}}=\min\left\{1,\frac{D_{k,a}}{\sum_{l=1}^{L}\tr{\bR_{k,a}\bQ^{\mathrm{0}}_{{k,l}}}}\right\},\quad \forall k,a.
	\end{align}
	
	\item \emph{Worst-case Power Backoff} \cite{YLH15}: Similar to \eqref{equ:Adpbackoff3}, we introduce the backoff factor as 
	\begin{align}\label{equ:WCbackoff1}
	\alpha^{\mathrm{wst}}=\min\left\{1,\frac{D_{k,a}}{\mathrm{SAR^{wst}}}\right\}, \quad \forall k,a,
	\end{align}
	where
	\begin{align}\label{equ:WCbackoff2}
	\mathrm{SAR^{wst}}=\underset{k,a}\max \underset{\sum_{l=1}^{L}\tr{\bQ_{k,l}}\leq\Pmaxk}\max\  \sum\limits_{l=1}^{L}\tr{\bR_{k,a}\bQ_{{k,l}}}.
	\end{align}
	Then, the worst-case power backoff approach makes the final result of the problem $\eqref{eq:problemlow}$ without \eqref{plowc} meet the SAR constraints by reducing the power budgets as 
	\begin{align}\label{equ:WCbackoff3}
	\bQ^{\mathrm{wc}}_{\mathrm{opt}}=\argmax{\sum_{l=1}^{L}\tr{\bQ_{k,l}}\leq\alpha^{\mathrm{wst}}\Pmaxk}\ \mathrm{EE}(\bQ,\bpi), \quad \forall k.
	\end{align}
\end{itemize}

\figref{fig:eepbackoff} shows that our proposed EM exposure-aware EE optimization design can achieve better EE performance compared with the baseline approaches.\footnote{Note that the complexity of both adaptive power backoff and worst-case power backoff can be estimated as $\mathcal{O}\left((I_4I_3+1)I_2\left(I_1\sum_{k}LN_k^3+K^x\right)\right)$. When the total number of SAR constraints is small, the complexity of our proposed method has the same order as that of backoff methods.} It can be observed that these approaches achieve the same EE when $\Pmax$ is low. It is consistent with the results shown in \figref{fig:eeDdifp3} that the power constraints are the main limits of system EE in the region of small power budgets, where SAR constraints can be naturally satisfied under this condition. Then, the backoff factors $\alpha_{k}^{\mathrm{adp}}, \forall k$ and $\alpha^{\mathrm{wst}}$ are equal to one, which means the power only constrained problem provides the same solutions of $\eqref{eq:problemlow}$. In addition, \figref{fig:eepbackoff} indicates that SAR constraints restrict the system EE in a different way from power constraints. Actually, SAR constraints consider both the amplitude and the phase differences between any two transmit antennas, while the power constraints only restrict the amplitude of transmit signals. Since the backoff approaches regard the SAR constraint as the additional power constraint in the overall optimization, the corresponding optimization results are inferior to that of our proposed approach.

\section{Conclusion}\label{sec:conclusion}
To summarize, we investigated the transmission design in the criterion of EE maximization with EM exposure constraints for uplink RSMA communications. Specifically, we optimized the transmit covariance matrices and decoding order in the non-convex mixed integer program, which is basically difficult to handle. We first applied the DE method to obtain an approximate expression of the ergodic EE for the reduction of optimization complexity. Later, we proposed a modified water-filling scheme to obtain sub-optimal solutions of covariance matrices with a given decoding order, where MM and Dinkelbach's methods were adopted to iron out the difficulty caused by the non-convexity. Then, we design the decoding order on its feasible set by using a greedy approach, which can reduce the overall complexity compared with the exhaustive approach. In the simulations, we confirmed the effectiveness of the overall algorithm, which converges quickly within a few iterations. In addition, the impact of the EM exposure constraints and the number of layers split from each user on the EE performance were presented for the uplink RSMA transmission. The numerical results verified the EE performance gain of uplink RSMA compared with the NOMA, SDMA, FDMA, and TDMA schemes. Since we actively considered the EM exposure constraints in the transmission design, our proposed EM exposure-aware water-filling scheme was shown to have superior EE performance compared with several baselines such as the adaptive and worst-case power backoff approaches.

\appendices

\section{Proof of \propref{Prop:Water-Filling}}\label{app:B}
As indicated in \cite{LGX16}, the DE of \eqref{equ:rkl+} is an asymptotic approximation function generated by the iterations of DE parameters, which has been concluded that
\begin{align}
\frac{\partial R_{k,l}^+(\bQ)}{\partial [\bTheta_k(\wbPhi_{k,l}^{-1}\bK_{k,l}^{-1})]_{j,j}}=\frac{\partial R_{k,l}^+(\bQ)}{\partial [\wbGamma_{k,l}\bK_{k,l}^{-1}]_{i,i}}=0,\ntb \forall i=1,...,M,\ j=1,...,N_k.
\end{align}
Accordingly, the DE can be regarded as a function of $\bQ_{k,l}$ with fixed DE parameters. The derivative of $R_{k,l}^+(\bQ)$ in \eqref{equ:DERkl} over $\bQ_{k,l}$ is given by \cite[Th. 4]{LGX16}
\begin{align}\label{equ:Derivative}
\frac{\partial R_{k,l}^+(\bQ)}{\partial \bQ_{k,l}}=(\bI_{N_k}+\bGamma_{k,l}\bQ_{k,l})^{-1}\bGamma_{k,l}.
\end{align}
Then, the optimal $\bQ$ and dual variables satisfy the Karush-Kuhn-Tucker (KKT) conditions:
\begin{align}
&\frac{\partial\mathcal{LA}}{\partial\bQ_{k,l}}=(\bI_{N_k}+\bGamma_{k,l}\bQ_{k,l})^{-1}\bGamma_{k,l}-\bS_{k,l}=\mathbf{0},\ \forall k,l,\label{equ:KKT}\\
&\mu_k\left(\tr{\bQ_{k,l}}-\Pmaxk\right)=0,\ \forall k,\label{equ:KKT2}\\ &\lambda_{k,a}\left(\tr{\bR_{k,a}\bQ_{k,l}}-D_{k,a}\right)=0,\ \forall k,a,\label{equ:KKT3}
\end{align}
where
\begin{align}\label{equ:Skl2}
\bS_{k,l}=&\sum\limits_{(\bar{p},\bar{q})\in\bar{\mathcal{\bm{Q}}}_{k,l}}(\bDelta_{k,l}^{\bar{p},\bar{q}})^{(\ell)}+(\eta^{(\ell),[t]}\xi_k+\mu_k)\bI_{N_k}+\ntb
&\sum\limits_{a=1}^A\lambda_{k,a}\bR_{k,a}.
\end{align}
With given $\bGamma_{k,l}$ and feasible dual variables $\{\mu_k\}_{\forall k}$ and $\{\lambda_{k,a}\}_{\forall k,a}$, the inner optimization of \eqref{equ:dual_problem} in terms of the optimization variable $\bQ_{k,l}$ is equivalent to the problem with the same KKT condition, which can be expressed as
\begin{align}\label{eqKKT}
\underset{\bQ_{k,l}\succeq\mathbf{0}}\max \quad \log\det\left(\bI_{N_k}+\bGamma_{k,l}\bQ_{k,l}\right)-\tr{\bS_{k,l}\bQ_{k,l}}.
\end{align}
Therefore, when the optimal dual variables $\{\mu_k^{\star}\}_{\forall k}$ and $\{\lambda_{k,a}^{\star}\}_{\forall k,a}$ are obtained by the outer optimization of \eqref{equ:dual_problem}, the solution of problem \eqref{eqKKT} are the same as \eqref{eq:problem5} in terms of the optimization variable $\bQ_{k,l}$.
Following the proof of \cite[Th. 3.6]{YLH15}, the analytical solutions of problem \eqref{eqKKT} can be derived as the form of \eqref{equ:wfsol}, which is omitted in this paper. This concludes the proof.

\bibliographystyle{IEEEtran}
\bibliography{Refabrv,EMC_RSMA_ref}

\end{document}